# Partitioning Patches into Test-equivalence Classes for Scaling Program Repair

Sergey Mechtaev
National University of Singapore
mechtaev@comp.nus.edu.sg

Xiang Gao
National University of Singapore
gaoxiang@comp.nus.edu.sg

Shin Hwei Tan
National University of Singapore
shinhwei@comp.nus.edu.sg

Abhik Roychoudhury
National University of Singapore
abhik@comp.nus.edu.sg

## ABSTRACT

Automated program repair is a problem of finding a transformation (called a patch) of a given incorrect program that eliminates the observable failures. It has important applications such as providing debugging aids, automatically grading assignments and patching security vulnerabilities. A common challenge faced by all existing repair techniques is scalability to large patch spaces, since there are many candidate patches that these techniques explicitly or implicitly consider.

The correctness criterion for program repair is often given as a suite of tests, since a formal specification of the intended program behavior may not be available. Current repair techniques do not scale due to the large number of test executions performed by the underlying search algorithms. We address this problem by introducing a methodology of patch generation based on a test-equivalence relation (if two programs are "test-equivalent" for a given test, they produce indistinguishable results on this test). We propose two test-equivalence relations based on runtime values and dependencies respectively and present an algorithm that performs on-the-fly partitioning of patches into test-equivalence classes.

Our experiments on real-world programs reveal that the proposed methodology drastically reduces the number of test executions and therefore provides an order of magnitude efficiency improvement over existing repair techniques, without sacrificing patch quality.

## 1 INTRODUCTION

As every developer knows, debugging is difficult and extremely time-consuming. Due to the slow adoption of automated verification and debugging techniques, finding and eliminating defects remains mostly a manual process. Automated patch generation approaches can potentially alleviate this problem since they have been shown to be able to address defects in real-world programs and require minimal developer involvement. Specifically, they have been successfully applied for providing debugging hints [Tao et al. 2014], automatically grading assignments [Rolim et al. 2017; Yi et al. 2017] and patching security vulnerabilities [Mechtaev et al. 2016]. However, the problem of huge search spaces pose serious challenges for current program repair techniques.

The goal of program repair is to modify a given incorrect program to eliminate the observable failures. Specifically, the goal of *test-driven* program repair is to modify the buggy program so that it passes all given tests. Patches (program modifications) that pass all given tests are referred to as *plausible* in the program repair literature [Qi et al. 2015]. Since a test-suite is an incomplete specification, plausible patches may not coincide with user's intentions but may merely *overfit* the tests [Smith et al. 2015]. To address this problem, existing techniques define a cost function (priority) on the space of candidate patches and search for a patch that optimizes this function. For example, changes can be prioritized based on syntactic distance [Mechtaev et al. 2015], semantic distance [D'Antoni et al. 2016] and information learned from human patches [Long and Rinard 2016b].

Patch generation systems need to consider large spaces of possible modifications in order to address many kinds of defects. One of the key challenges of program repair is scalability to large search spaces. Current techniques may require substantial time to generate patches and yet they consider and generate only relatively simple program transformations [Long and Rinard 2015]. This impacts the ability of program repair to produce human-like repairs, since human patches often involve complex source code modifications. Besides that, a recent study [Long and Rinard 2016a] demonstrated that extending the search space with more transformations may cause repair systems to find fewer correct repairs because of the increased search time.

Although existing test-driven program repair techniques employ different methodologies (e.g. GenProg [Weimer et al. 2013] uses genetic programming, SemFix [Nguyen et al. 2013] and Angelix [Mechtaev et al. 2016] are based on constraint solving, Prophet [Long and Rinard 2016b] is based on machine learning), they all search for plausible patches by repeatedly executing tests. For instance, GenProg explicitly generates and tests syntactic changes, Angelix explores deviations of execution paths for given tests using symbolic execution (which can be considered as a variant of test execution), Prophet applies machine learning to rank patches that are also validated through test executions. Due to the high cost of test executions in large real-world programs, the number of performed test executions is the main bottleneck of many existing program repair algorithms.

*This work.* The purpose of this work is to improve the scalability of program repair without sacrificing the quality of generated patches. In order to achieve this, we propose a methodology based on a test-equivalence relation [Just et al. 2014; Le et al. 2014]. If two programs are test-equivalent for a test, then the programs produce indistinguishable results on that test:

*Definition 1.1 (Test-equivalence).* Let $\mathcal{P}$ be a set of programs, $t$ be a test. An equivalence relation (reflexive, symmetric and transitive) $\overset{t}{\sim} \subset \mathcal{P} \times \mathcal{P}$ is a *test-equivalence relation* for $t$ if it is consistent with the results of $t$, that is $\forall p_1, p_2 \in \mathcal{P}$, if $p_1 \overset{t}{\sim} p_2$ then $p_1$ and $p_2$ either both pass $t$ or both fail $t$.

Identifying test-equivalence classes of patches enables our algorithm to reduce the number of required test executions since a single execution is sufficient to evaluate all patches in the same test-equivalence class.

*Contributions.* The main contributions of this work are described in the following.

(1) We propose the use of test-equivalence relations to drastically prune the search space explored for the purpose of program repair.
(2) We define two test-equivalence relations (based on runtime values and dependencies) for spaces of program modifications generated through program synthesis. These two test-equivalence relations can be composed, thereby producing a more effective (coarse-grained) partitioning of patches into test-equivalence classes.
(3) We introduce a new patch space exploration algorithm that performs on-the-fly (during test execution) partitioning of patches into test-equivalence classes, thereby achieving efficient program repair that requires fewer test executions to generate a patch.
(4) We conduct an evaluation of the algorithm on real-world programs from the GenProg ICSE'12 benchmark [Le Goues et al. 2012a]; it demonstrates that test-equivalence significantly reduces the number of required test executions and therefore increases the efficiency of test-driven program repair and scales it to larger search spaces without sacrificing patch quality.

*Outline.* In the next section, we provide examples demonstrating limitations of existing techniques and formulate key insights of our methodology. Section 3 formally defines the two test-equivalence relations. Section 4 introduces a repair algorithm based on these relations, Section 5 describes its implementation and Section 6 presents its experimental evaluation. Section 7 discusses related work, Section 8 discusses future research directions and Section 9 concludes.

```
...
(*tif->tif_close)(tif);
if (tif->tif_rawcc > 0
    && tif->tif_rawcc != orig_rawcc
    && (tif->tif_flags & TIFF_BEENWRITING)!= 0
    && !TIFFFlushData1(tif)) {
      TIFFErrorExt(tif->tif_clientdata,
        module,
        "Error␣flushing␣data␣before␣directory␣write
            ");
    return (0);
  }
...
```

**(a) Incorrect condition in Libtiff (rev. 0661f81).**

```
...
(*tif->tif_close)(tif);
if (tif->tif_rawcc > 0
    && (tif->tif_flags & TIFF_BEENWRITING)!= 0
    && !TIFFFlushData1(tif)) {
      TIFFErrorExt(tif->tif_clientdata,
        module,
        "Error␣flushing␣data␣before␣directory␣write
            ");
    return (0);
  }
...
```

**(b) Developer patch for incorrect condition.**

**Figure 1: Defect in Libtiff library from GenProg ICSE'12 benchmark.**

## 2 MOTIVATING EXAMPLES

This section gives three examples demonstrating limitations of existing techniques: a large number of redundant test executions, ineffectiveness in searching for optimal repairs and restricted applicability. We also formulate key insights that enable our method to address these limitations.

### 2.1 Example: repairing conditions

Consider a defect in the revision 0661f81 of Libtiff[1] from the GenProg ICSE'12 benchmark. The code in Figure 1a is responsible for flushing data written by the compression algorithm, and the defect is caused by the wrong highlighted condition. Libtiff test-suite contains 78 tests, and this defect is manifested by a failing test called "tiffcp-split". Figure 1b demonstrates the developer patch that modifies the wrong condition by removing the clause `tif->tif_rawcc != orig_rawcc`.

We demonstrate how existing automated program repair algorithms generate a patch for this condition. First, repair algorithms perform fault localization to identify suspicious program statements. The number of localized statements in existing tools may vary from tens to thousands depending on algorithms and configurations (it can potentially include all executed statements). In this example, we consider only the location of the buggy expression highlighted in Figure 1a.

Second, program repair algorithms define a *search space* of candidate patches. In this work, we primarily focus on two state-of-the-art approaches that have been shown to scale to large real-world programs: Angelix [Mechtaev et al. 2016] and Prophet [Long and Rinard 2016b]. Specifically, our goal was to support a combination of transformations implemented in these systems. Thus, the search space for the highlighted condition includes all possible replacements of its subexpressions by expressions constructed from visible program variables and C operators, refinements (e.g. appending `&& EXPR` and `|| EXPR`), replacements of operators and swapping arguments. In total, the search space in our synthesizer contains 56 243 modifications of the condition.

Finally, program repair algorithms explore the search space in order to try to find a modification that passes all given tests. Existing search space exploration methods can be classified into syntax-based and semantics-based. *Syntax-based* algorithms explicitly generate and test syntactic changes. In this example, a syntax-based algorithm have to execute the failing test 56 243 times to evaluate all candidates[2]. Since there are 78 tests in the test-suite, 907 457 test executions are required to explore the search space[3] (we say that an element of a search space is *explored* if the algorithm identifies if it passes all the tests or fails at least one). Given the high cost of test execution, this approach has poor scalability.

[1]Libtiff is a software library that provides support for TIFF image format: http://simplesystems.org/libtiff/

[2]Since the search space contains the correct patch in this example, the algorithm can stop search earlier after the patch is found. Then, the number of test executions depends on the exploration order.

[3]This data is obtained by executing our implementation of syntactic enumeration.

```
1. ((tif->tif_rawcc > 0) && (tif->tif_rawcc != orig_rawcc))
    || (tif->tif_flags & TIFF_BEENWRITING)
2. ((tif->tif_rawcc > 0) || (tif->tif_rawcc != orig_rawcc))
    && (tif->tif_flags & TIFF_BEENWRITING)
3. ((tif->tif_rawcc == 0) && (tif->tif_rawcc != orig_rawcc))
    && (tif->tif_flags & TIFF_BEENWRITING)
4. (((tif->tif_rawcc > 0) && (tif->tif_rawcc != orig_rawcc))
    && (tif->tif_flags & TIFF_BEENWRITING)) || (imagedone >= orig_rawcc)
5. (((tif->tif_rawcc > 0) && (tif->tif_rawcc != orig_rawcc))
    && (tif->tif_flags & TIFF_BEENWRITING)) || (tif->tif_flags >= 74)
```

**Figure 2: Each of 56 243 search space elements is test-equivalent to one of these 5 expressions.**

*Semantics-based* techniques (e.g. Semfix [Nguyen et al. 2013], SPR [Long and Rinard 2015], Angelix and Prophet) split exploration into two phases. First, they infer a synthesis specification for the identified expression through path exploration. For this example, they enumerate and execute sequences of condition values (e.g. *true*, *true*, *true*, *false*, ...) to find those sequences that enable the program to pass the test. Second, they synthesize a modification of the condition to match the inferred specification. In this example, there are 256 possible execution paths (the condition is evaluated multiple times during the test execution), therefore a semantics-based algorithm performs 256 test executions for the failing tests, and 1320 for the whole test-suite[4]. Although semantics-based techniques were shown to be more scalable [Long and Rinard 2015], they are subject to the *path explosion problem*: the number of execution paths can be infinite. To address this, current systems introduce a bound for the number of explored paths, however it may affect their effectiveness: if a path followed by the correct patch is omitted, then this correct patch cannot be generated.

The algorithm proposed in this work performs on-the-fly partitioning of program modifications into test-equivalence classes. We demonstrate the effect of the relation $\overset{t}{\sim}_{value}$ described in Section 3.3. Two modifications of a program expression are test-equivalent w.r.t. $\overset{t}{\sim}_{value}$ if they are evaluated into the same sequences of values during the test execution. Surprisingly, the space of 56 243 modifications can be partitioning into only 5 test-equivalence classes for the failing test "tiffcp-split" w.r.t. $\overset{t}{\sim}_{value}$; five elements of the search space that represent different test-equivalence classes are given in Figure 2. Since all patches in the same test-equivalence class exhibit the same behaviour for the test, the failing test can be executed only 5 times to evaluate all candidates.

[4]This data is obtained by executing Angelix.

Our algorithm computes test-equivalence classes for each test in the test-suite. However, since test-equivalence classes for different tests may intersect, our algorithm takes advantage of this to skip redundant execution across different tests. Specifically, for each next test it only evaluates subspaces of modifications that are not included into failing test-equivalence classes of previously executed tests. Meanwhile, semantics-based techniques perform specification inference for each test independently without reusing information across tests. As a result, our algorithm requires only 103 test executions to evaluate all 56 243 modifications with the whole test-suite.

*Key insight.* The key insight that enables our method to reduce the number of required test executions is that, compared with techniques that explore execution paths, it takes the expressiveness of the patch space into account (e.g. it identifies that only 5 out of 256 possible execution paths are induced by the considered set of 56 243 transformations). Compared with syntactic enumeration, it substantially reduces executions since a single execution evaluates a whole test-equivalence class.

## 2.2 Example: optimal repair

Since a test-suite is an incomplete specification, test-driven program repair suffers from the test overfitting problem [Smith et al. 2015]. To address this issue, state-of-the-art techniques define a priority (a cost function) in the space of patches and search for a program modification that optimizes this function. Ideally, this function should assign higher cost to

```
while i > 0 do
  if * then
    c := c + 1
  fi;
  i := i - 1
od
```

**(a) Buggy program $p$.**

$$\begin{aligned}\mathcal{P} \coloneqq \{\ & p[{}^{*}/_{\mathtt{i} \geq 0}],\\ & p[{}^{*}/_{\mathtt{c} \geq 0}],\\ & p[{}^{*}/_{\mathtt{i}\ \mathrm{mod}\ 2\ =\ 1}],\\ & p[{}^{*}/_{\mathtt{i}\ \mathrm{mod}\ 2\ =\ 0}],\\ & p[{}^{*}/_{\mathtt{i} > 1}]\ \}\end{aligned}$$

**(b) Search space.**

$$\begin{aligned}\kappa(p[{}^{*}/_{\mathtt{i} \geq 0}]) &\coloneqq 0.1\\ \kappa(p[{}^{*}/_{\mathtt{c} \geq 0}]) &\coloneqq 0.2\\ \kappa(p[{}^{*}/_{\mathtt{i}\ \mathrm{mod}\ 2\ =\ 1}]) &\coloneqq 0.3\\ \kappa(p[{}^{*}/_{\mathtt{i}\ \mathrm{mod}\ 2\ =\ 0}]) &\coloneqq 0.4\\ \kappa(p[{}^{*}/_{\mathtt{i} > 2}]) &\coloneqq 0.5\end{aligned}$$

**(c) Cost function.**

**Figure 3: Example of optimal program repair problem.**

overfitting patches. For instance, Prophet [Long and Rinard 2016b] shown how such a cost function learned from human patches enables the generation of more correct repairs.

Consider a program $p$ in Figure 3a that counts odd numbers in the interval $(0, i]$. The $*$ indicates a wrong condition that has to be modified by the repair algorithm (the correct condition is $\mathtt{i}\ \mathrm{mod}\ 2 = 1$). We denote a program obtained by substituting $*$ with an expression $e$ as $p[{}^{*}/_{e}]$. The repair algorithm searches for a plausible patch (a substitution of $*$ with a condition) from the space $\mathcal{P}$ in Figure 3b such that the resulting program passes the test $t$ defined as follows:

$$t \coloneqq (\{\ i \mapsto 4,\ c \mapsto 0\ \},\ \lambda\sigma.\ \sigma(c) = 2)$$

where $t$ is pair of (1) an initial program state (mapping from variables to values) and (2) a test assertion (a boolean function over program states) denoted using lambda notation. We assume that $*$ is such that $p$ fails $t$. Besides that, we consider a cost function $\kappa$ defined for the considered space of substitutions in Figure 3c. The goal is to find a plausible patch with the lowest cost.

In order to find a patch for the example program, techniques like Angelix and Prophet enumerate possible sequences of values that a condition can take during test execution. Since there can be potentially infinite number of such sequences, existing approaches introduce a bound for the number of explored sequences and use an exploration heuristics to choose which sequences to explore. For instance, Prophet enumerates sequences where the condition first always takes the true branch until a certain point after which it always takes the false branch. Thus, for the considered example it would enumerate the following sequences:

$$\{\ \mathit{true}, \mathit{true}, \mathit{true}, \mathit{true}\ \},\ \{\ \mathit{true}, \mathit{true}, \mathit{true}, \mathit{false}\ \},$$
$$\{\ \mathit{true}, \mathit{true}, \mathit{false}, \mathit{false}\ \},\ \{\ \mathit{true}, \mathit{false}, \mathit{false}, \mathit{false}\ \}.$$

For each of these sequences, Prophet executes the program with the test $t$ in such a way that the condition $*$ takes the values as in this sequence during the execution. Only the third sequence $\{\ \mathit{true}, \mathit{true}, \mathit{false}, \mathit{false}\ \}$ enables the program to pass $t$, therefore it will be selected as a specification for expression synthesis. The synthesizer will find the expression $\mathtt{i} > 2$ obtaining a suboptimal patch $p[{}^{*}/_{\mathtt{i} > 2}]$ with the cost 0.5, since this is the only expression from the search space satisfying the specification. However, the correct expression $\mathtt{i}\ \mathrm{mod}\ 2 = 1$ with a lower cost 0.3 cannot be generated, since the corresponding sequence $\{\ \mathit{false}, \mathit{true}, \mathit{false}, \mathit{true}\ \}$ is not explored by the algorithm.

In contrast to techniques like Angelix and Prophet, our algorithm iterates through the search space in such a way that at each steps it selects and evaluates an unevaluated candidate with the lowest cost. Specifically, it starts by choosing the candidate $p[{}^{*}/_{\mathtt{i} \geq 0}]$ with the cost 0.1. It executes this candidate on-the-fly computing its test-equivalence class w.r.t. $\overset{t}{\sim}_{value}$ described in Section 2.1. This class contains the program $p[{}^{*}/_{\mathtt{c} \geq 0}]$, since the conditions $\mathtt{i} \geq 0$ and $\mathtt{c} \geq 0$ produce the same sequence of values $\{\ \mathit{true}, \mathit{true}, \mathit{true}, \mathit{true}\ \}$ for $t$. Since $p[{}^{*}/_{\mathtt{i} \geq 0}]$ does not pass the test, the whole corresponding test-equivalence class is marked as failing. Next, it selects $p[{}^{*}/_{\mathtt{i}\ \mathrm{mod}\ 2\ =\ 1}]$ with the cost 0.3 since $p[{}^{*}/_{\mathtt{c} \geq 0}]$ was indirectly evaluated through test-equivalence at the previous step. Since this candidate passes the test, the algorithm outputs it as a found repair.

```
...
clear_bufs();
to_stdout = 1;
part_nb = 0;
ifd = part_nb;
if (decompress) {
  method=get_method(ifd);
...
```

**(a) Before if-statement.**

```
...
clear_bufs();
to_stdout = 1;
part_nb = 0;
ifd = 0;
if (decompress) {
  method=get_method(ifd);
...
```

**(b) Before if-statement.**

```
...
clear_bufs();
to_stdout = 1;
part_nb = 0;
if (decompress) {
  ifd = part_nb;
  method=get_method(ifd);
...
```

**(c) Inside if-statement.**

**Figure 4: Candidate patches for defect of Gzip from GenProg ICSE'12 benchmark.**

*Key insight.* Our algorithm guides exploration based on a given cost function and focuses on high priority areas of the space of patches. By construction, if it finds a patch, then this patch is guaranteed to be the global optimum in the search space w.r.t. the cost function. Angelix and Prophet, on the other hand, may spend executions for value sequences that correspond to suboptimal candidates or correspond to no candidates at all (e.g. { *false*, *true*, *true*, *true* }), and therefore may miss the best patch in their search space.

## 2.3 Example: repairing assignments

Although current program repair approaches has been shown to be relatively effective in modifying existing program expressions, they provide limited support for more complex transformations. In this work, we consider one such transformation that inserts a new assignment statement to the buggy program. Techniques like Prophet and GenProg can generate patches by copying/moving existing program assignments, however this approach has limitations: (1) assignments for local variables cannot be copied from different parts of the program because of their scope and (2) each insertion of an assignment is validated separately, which yield a large number of required test executions. Existing techniques do not apply specification inference for assignment synthesis (as described in Section 2.1 for conditions) because such specification has to encode all possible side effects that can be caused by assignment insertion (for each variable that can appear in the left-hand side of the assignment), which makes inferring such specification infeasible for large programs.

We show how test-equivalence can scale assignment synthesis for a defect in Gzip from the GenProg ICSE'12 benchmark[5]. Consider three candidate patches in Figure 4 that insert the highlighted statements at several program locations. First, our algorithm identifies that the program in Figure 4a is test-equivalent to the program in Figure 4b (w.r.t. the relation $\overset{t}{\sim}_{value}$ described previously in Section 2.1) since they differ only in the right-hand side of the highlighted assignments and the corresponding expressions take the same values during test execution. Second, using a simple dynamic dependency analysis our algorithm identifies that the program in Figure 4a is test-equivalent to the program in Figure 4c since (1) they insert the same assignment at different program locations, (2) both these locations are executed by the test since the true branch of the if-statement is taken during the test execution and (3) the variables `ifd` and `part_nb` are not used/modified between these locations during test execution. We refer to such a test-equivalence relation as $\overset{t}{\sim}_{deps}$. Finally, our algorithm merges the results of the two analyses (as the transitive closure of their union) and determines that the program in Figure 4b is test-equivalent to the program in Figure 4c. Therefore, a single test execution is sufficient to evaluate all these patches.

*Key insight.* Since test-equivalence is a weaker property than the property of "passing the test" expressed by the inferred specification in semantics-based techniques, it permits using more lightweight analysis techniques. Specifically, we demonstrate that a composition of two lightweight test-equivalence analyses enables us to scale assignment synthesis.

[5]Gzip is a file compression/decompression application: `https://www.gnu.org/software/gzip/`

$\langle Stmt\rangle$ ::= $\langle Var\rangle$ := $\langle AExpr\rangle$
  | `skip`
  | $\langle Stmt\rangle$ ; $\langle Stmt\rangle$
  | `if` $\langle BExpr\rangle$ `then` $\langle Stmt\rangle$ `else` $\langle Stmt\rangle$ `fi`
  | `while` $\langle BExpr\rangle$ `do` $\langle Stmt\rangle$ `od`

$\langle AExpr\rangle$ ::= $\langle Var\rangle$
  | $\langle Num\rangle$
  | $\langle AExpr\rangle$ $\langle AOp\rangle$ $\langle AExpr\rangle$

$\langle BExpr\rangle$ ::= `true`
  | `false`
  | $\langle BExpr\rangle$ $\langle BOp\rangle$ $\langle BExpr\rangle$
  | $\langle AExpr\rangle$ $\langle ROp\rangle$ $\langle AExpr\rangle$

$\langle AOp\rangle$ ::= `+` | `-` | `*` | ...

$\langle BOp\rangle$ ::= `and` | `or` | ...

$\langle ROp\rangle$ ::= `<` | `<=` | `=` | ...

**Figure 5: Syntax of programming language $\mathcal{L}$.**

# 3 TEST-EQUIVALENCE RELATIONS

This section formally introduces two test-equivalence relations for spaces of program modifications generated through program synthesis. In the subsequent Section 4, we demonstrate how these relations can be applied for scaling patch generation. However, we believe that these relations can be also used in different domains; other potential applications are discussed in Section 7.

## 3.1 Preliminaries

We introduce our methodology for an imperative programming language $\mathcal{L}$. The syntax of $\mathcal{L}$ is defined in Figure 5, where $\mathbb{Z}$ is the integer domain, $\mathbb{B}$ is the boolean domain (true and false), *Stmt* is a set of statements, *AExpr* is a set of arithmetic expressions, *BExpr* is a set of boolean expressions, $Expr = AExpr \cup BExpr$, *Num* is a set of integer literals, *Var* is a set of variables over $\mathbb{Z}$. A program in $\mathcal{L}$ is a sequence of statements. The semantics of $\mathcal{L}$ is defined in Figure 6, where program state $\sigma : Var \rightarrow \mathbb{Z}$ is a function from program variables into values, $\Sigma$ is a set of program states. We indicate a modification of a program state $\sigma$ where the value of the variable $v$ is updated to $n$ as $\sigma[v \mapsto n]$. We denote subsets of $\mathcal{L}$ as $\mathcal{P}$, subsets of expressions *Expr* as $\mathcal{E}$, all variables from *Var* encountered in an expression $e$ as $Var(e)$. We denote a program obtained by substituting a statement (expression) $s$ with a statement (expression) $s'$ in a program $p$ as $p[^{s}/_{s'}]$.

*Definition 3.1 (Test).* Let $p \in \mathcal{L}$ be a program, $t \in \Sigma \times (\Sigma \rightarrow \mathbb{B})$ be a test, that is a pair $(\sigma_{in}, \phi)$, where $\sigma_{in}$ in the initial program state (input) and $\phi$ is the test assertion (a boolean function over program states). We say that $p$ passes $t$ (indicated as $Pass[p, t]$) iff $\langle p, \sigma_{in}\rangle \Downarrow \sigma_{out} \wedge \phi(\sigma_{out})$.

## 3.2 Generalized synthesis

We define *synthesis specification* as a finite set of input-output pairs (an input is represented by a program state from $\Sigma$ and an output is an integer or a boolean value); we denote the set of all specifications as $Spec := 2^{\Sigma\times(\mathbb{Z}\cup\mathbb{B})}$.

*Definition 3.2 (Synthesis procedure).* A syntax-guided synthesis procedure $synthesize : 2^{Expr} \times Spec \rightarrow Expr$ is a function that takes a set of expressions (the synthesis search space) and a specification, and returns an expression from the search space that meets the specification. Specifically, for a given search space $\mathcal{E}$ and specification *spec*

$$\text{if } synthesize(\mathcal{E}, spec) = e \text{ then } e \in \mathcal{E} \wedge \bigwedge_{\sigma, n \in spec} \langle e, \sigma\rangle \Downarrow n.$$

In order to integrate synthesis with test-equivalence analysis, we impose additional requirements for the program synthesizer: it should define a value-projection operator over its search space.

The *value-projection operator* $\Pi^{value}_{\sigma,n}$ produces a maximal subset of a given set of expressions consisting only of expressions that are evaluated into $n$ in the context $\sigma$:

$$\Pi^{value}_{\sigma,n}(\mathcal{E}) = \{e \mid e \in \mathcal{E} \wedge \langle e, \sigma\rangle \Downarrow n\}$$

In this work, we use an enumerative synthesizer [Alur et al. 2013] that demonstrated positive results in program synthesis competitions[6]. Since it represents the search space explicitly as a set of expressions, it is straightforward to realize the value-projection operator in such a synthesizer. Other possible realizations are discussed below.

*Enumerative synthesis.* In enumerative synthesis [Alur et al. 2013], the search space is represented explicitly as

[6]SyGuS-Comp 2014: `http://www.sygus.org/SyGuS-COMP2014.html`

$$\text{VAR}\quad \frac{}{\langle v,\sigma\rangle \Downarrow \sigma(v)} \qquad \text{NUM}\quad \frac{}{\langle n,\sigma\rangle \Downarrow n} \qquad \text{OP}\quad \frac{\langle e_1,\sigma\rangle \Downarrow n_1 \quad \langle e_2,\sigma\rangle \Downarrow n_2 \quad n_3 = n_1\ op\ n_2}{\langle e_1\ op\ e_2,\sigma\rangle \Downarrow n_3} \qquad \text{ASSIGN}\quad \frac{\langle e,\sigma\rangle \Downarrow n}{\langle v := e,\sigma\rangle \Downarrow \sigma[v \mapsto n]}$$

$$\text{SEQ}\quad \frac{\langle s_1,\sigma\rangle \Downarrow \sigma_1 \quad \langle s_2,\sigma_1\rangle \Downarrow \sigma_2}{\langle s_1\ ;\ s_2,\sigma\rangle \Downarrow \sigma_2} \qquad \text{IF-TRUE}\quad \frac{\langle e,\sigma\rangle \Downarrow true \quad \langle s_1,\sigma\rangle \Downarrow \sigma_1}{\langle \texttt{if}\ e\ \texttt{then}\ s_1\ \texttt{else}\ s_2\ \texttt{fi},\sigma\rangle \Downarrow \sigma_1} \qquad \text{IF-FALSE}\quad \frac{\langle e,\sigma\rangle \Downarrow false \quad \langle s_2,\sigma\rangle \Downarrow \sigma_2}{\langle \texttt{if}\ e\ \texttt{then}\ s_1\ \texttt{else}\ s_2\ \texttt{fi},\sigma\rangle \Downarrow \sigma_2}$$

$$\text{WHILE-TRUE}\quad \frac{\langle e,\sigma\rangle \Downarrow true \quad \langle s_1,\sigma\rangle \Downarrow \sigma_1 \quad \langle \texttt{while}\ e\ \texttt{do}\ s\ \texttt{od},\sigma_1\rangle \Downarrow \sigma_2}{\langle \texttt{while}\ e\ \texttt{do}\ s\ \texttt{od},\sigma\rangle \Downarrow \sigma_2} \qquad \text{WHILE-FALSE}\quad \frac{\langle e,\sigma\rangle \Downarrow false}{\langle \texttt{while}\ e\ \texttt{do}\ s\ \texttt{od},\sigma\rangle \Downarrow \sigma} \qquad \text{SKIP}\quad \frac{}{\langle \texttt{skip},\sigma\rangle \Downarrow \sigma}$$

**Figure 6: Semantics of $\mathcal{L}$. $v$ — variables, $n$ — integer values, $e$ — expressions, $s$ — statements, $\sigma$ — program states.**

**ALGORITHM 1:** Value-projection operator via enumerative synthesis

**Input**: set of expressions $\mathcal{E}$, program state $\sigma$, value $n$
**Output**: set of expressions $\mathcal{E}'$

$\mathcal{E}' := \emptyset$;
**foreach** $e \in \mathcal{E}$ **do**
  **if** $\langle e, \sigma\rangle \Downarrow n$ **then**
    $\mathcal{E}' := \mathcal{E}' \cup \{e\}$;
**return** $\mathcal{E}'$;

a set of expressions. The value-projection operator can be implemented as shown in Algorithm 1.

*Symbolic synthesis.* In SMT-based component-based synthesis [Jha et al. 2010], the search space is represented implicitly through logical constraints. To synthesize an expression from a given specification, it solves the following formula:

$$\exists e \in \mathcal{E}. \bigwedge_{\sigma, n \in spec} \langle e,\sigma\rangle \Downarrow n$$

To support quantification over expressions, it uses *location variables* (with prefix $l$) to encode all expressions constructed from given components $C$ (e.g. +, −, variables, etc) as follows.

$$\phi_{wpf} := \phi_{range} \wedge \phi_{cons} \wedge \phi_{acyc}$$

$$\phi_{range} := \bigwedge_{c\in C} \Big(0 \le lc^{out} < |C| \ \wedge \bigwedge_{k\in[1,NI(c)]} 0 \le lc_k^{in} < |C|\Big)$$

$$\phi_{cons} := \bigwedge_{(c,s)\in C\times C, c\ne s} lc^{out} \ne ls^{out}$$

$$\phi_{acyc} := \bigwedge_{c\in C, k\in[1,NI(c)]} lc^{out} > lc_k^{in}$$

$$\phi_{conn} := \bigwedge_{\substack{(c,s)\in C\times C\\ k\in[1..NI(s)]}} lc^{out} = ls_k^{in} \Rightarrow c^{out} = s_k^{in}$$

The algorithm also imposes library constraint ($\phi_{lib}$) that capture semantics of given components. For example, for the component $c := h_1 + h_2$, the library constraint is $c^{out} = c_1^{in} + c_1^{in}$. Given the above constraints, the search space is encoded through the formula $\phi$ defined as $\phi := \phi_{wpf} \wedge \phi_{lib} \wedge \phi_{conn}$. In this context, the value-projection operator can be implemented as follows:

$$\Pi^{value}_{\sigma,n}(\phi) := \phi \wedge e^{out} = n \wedge \bigwedge_{\text{variable } v\in C} v^{out} = \sigma(v)$$

where $e^{out}$ captures the output of the synthesized expression. Effectively, this operator conjoins the formula representing the search space with the input-output relation represented via $\sigma$ and $n$.

*DSL-based synthesis.* In FlashMeta [Polozov and Gulwani 2015] synthesis framework, the search space is compactly represented via version space algebra (VSA) [Mitchell 1982]. Assume that the set of expression $\mathcal{E}$ is defined through applications of operators to a given set of variables; we denote an operator as $F$. The grammar for a version space algebra $\widetilde{N}$ is defined as

$$\widetilde{N} := \{\ e_1, ..., e_k\ \} \mid U(\widetilde{N}_1, ..., \widetilde{N}_k) \mid F_{\bowtie}(\widetilde{N}_1, ..., \widetilde{N}_k)$$

such that

- $e \in \{\ e_1, ..., e_k\ \}$ if $\exists i.\ e = e_i$;
- $e \in U(\widetilde{N}_1, ..., \widetilde{N}_k)$ if $\exists i.\ e \in \widetilde{N}_i$;
- $e \in F_{\bowtie}(\widetilde{N}_1, ..., \widetilde{N}_k)$ if $e = F(e_1, ...e_n) \wedge \forall i.\ e_i \in \widetilde{N}_i$.

For version space algebra $\widetilde{N}$, FlashMeta implements a clustering operator denoted as $\widetilde{N}|_\sigma$. $\widetilde{N}|_\sigma$ is a mapping from values to version space algebras $\{\ n_1 \mapsto \widetilde{N}_1, ..., n_k \mapsto \widetilde{N}_k\ \}$ such that

- $\widetilde{N} = \widetilde{N}_1 \cup \ldots \cup \widetilde{N}$;
- $\widetilde{N}_i \cap \widetilde{N}_j = \emptyset$ for all $i \neq j$;
- $\forall e \in \widetilde{N}_i .\ \langle e, \sigma \rangle \Downarrow n_i$;
- $\forall i, j.\ i \neq j \rightarrow n_i \neq n_j$.

In this context, the value-projection operator can be implemented as follows:

$$\Pi^{value}_{\sigma,n}(\widetilde{N}) := \widetilde{N}|_{\sigma}(n)$$

### 3.3 Value-based test-equivalence relation

This section introduces a test-equivalence relation $\overset{t}{\sim}_{value}$ for spaces of programs that differ only in expressions. Intuitively, two programs $p$ and $p'$ such that $p' = p[^e/_{e'}]$ for some expressions $e$ and $e'$ are test-equivalent for some test $t$ if, during the executions of $p$ and $p'$ with $t$, the expressions $e$ and $e'$ are evaluated into the same values. An example of applying this relation is given in Section 2.1.

We define the relation $\overset{t}{\sim}_{value}$ *constructively* using an augmented semantics of $\mathcal{L}$. We chose this presentation since it simultaneously defines an algorithm of computing test-equivalence classes in spaces of modification generated via program synthesis. The implementation of this semantics via program instrumentation is discussed in Section 5.

The semantics in Figure 7 extends the semantics in Figure 6 by defining the function $\Downarrow_{value}$. It is parameterized by a predicate *Modified* : $Expr \rightarrow \mathbb{B}$ that marks the modified program expression, substitutions of which are analyzed for test-equivalence. The function $\Downarrow_{value}$ additionally maintains a set of expressions (denoted as $c$), such that the substitutions of the modified expression with $c$ form the computed test-equivalent class.

The augmented semantics describes an algorithm of identifying test-equivalence classes that, for a given set of expressions (the synthesis search space), "filters out" those that do not belong to the test-equivalent class of the current program by repeatedly applying the value-projection operator. The application of the value-projection operator to the current set of expressions $c$ is highlighted in Figure 7. Thus, it identifies all expressions from $c$ that produce the same value as the original expression at this evaluation step. Since each expression can be evaluated multiple times during test execution, the value-projection operator can also be applied multiple times. Therefore, the test-equivalence class is computed as $\Pi^{value}_{\sigma_1,n_1} \circ \Pi^{value}_{\sigma_2,n_2} \circ \ldots \circ \Pi^{value}_{\sigma_k,n_k}(\mathcal{E})$, where $\mathcal{E}$ is a set of all substitutions of the modified expression, $n_i$ are the values of the modified expression computed during test execution and $\sigma_i$ are the corresponding program states.

*Definition 3.3 (Value-based test-equivalence relation).* Let $\mathcal{P}$ be a set of programs, $t = (\sigma_{in}, \phi)$ be a test. $\overset{t}{\sim}_{value} \subset \mathcal{P} \times \mathcal{P}$ is a *value-based test-equivalence relation* iff $p_1 \overset{t}{\sim}_{value} p_2$ for $p_1, p_2 \in \mathcal{P}$ if $\exists e, e' \in Expr$ such that $p_2 = p_1[^e/_{e'}]$ and $\langle p_1, \sigma_{in}, \{e, e'\}\rangle \Downarrow_{value} \langle _, \{e, e'\}\rangle$ where $Modified := \lambda x.\ x = e$.

In this definition, we call two programs that differ only in expressions to be test-equivalent if the corresponding expressions produce the same values according to the semantics in Figure 6. Specifically, by passing the program $p_1$, the test input $\sigma_{in}$ and the set of expressions $\{e, e'\}$ as the arguments to $\Downarrow_{value}$, we obtain the same set $\{e, e'\}$ as the result.

PROPOSITION 3.4. *The relation $\overset{t}{\sim}_{value}$ is a test-equivalence relation according to Definition 1.1.*

The proposition above formally states that (1) $\overset{t}{\sim}_{value}$ is an equivalence relations and (2) if two programs that differ only in expressions are test-equivalent according to the semantics in Figure 7, then these two programs either both pass the test of both fail the test.

### 3.4 Dependency-based test-equivalence relation

This section introduces a test-equivalence relation $\overset{t}{\sim}_{deps}$ for spaces of programs that differ in locations in which an assignment statement is inserted. Let a *location* in program $p$ be a statement of $p$. We say that a program $p'$ is obtained by inserting the assignment $v := e$ at the location $l$ iff $p' = p[^l/_{v\,:=\,e;l}]$. Let $p$ be a program and programs $p_1$ and $p_2$ are obtained by inserting the assignment $v := e$ at the locations $l_1$ and $l_2$ of $p$ respectively. Informally, $p_1$ and $p_2$ are test-equivalent for some test $t$ if, during an execution of $p_1$ with $t$, (1) for each occurrence of $l_1$ in the execution trace there is a "matching" occurrence of $l_2$ (the variable $v$ is not read or overwritten between these occurrences and the variables $Var(e)$ are not overwritten between these occurrences), and (2) for each occurrence of $l_2$ in the execution trace there is a "matching"

$$\text{EXPR-MOD}\quad \frac{\mathit{Modified}(e) \qquad \langle e,\sigma\rangle \Downarrow n \qquad c' = \Pi^{value}_{\sigma,n}(c)}{\langle e,\sigma,c\rangle \Downarrow_{value} \langle n,c'\rangle}$$

$$\text{EXPR-NMOD}\quad \frac{\neg\mathit{Modified}(e) \qquad \langle e,\sigma\rangle \Downarrow n}{\langle e,\sigma,c\rangle \Downarrow_{value} \langle n,c\rangle} \qquad \text{ASSIGN}\quad \frac{\langle e,\sigma,c\rangle \Downarrow_{value} \langle n,c'\rangle}{\langle v := e,\sigma,c\rangle \Downarrow_{value} \langle \sigma[v \mapsto n],c'\rangle}$$

$$\text{IF-TRUE}\quad \frac{\langle e,\sigma,c\rangle \Downarrow_{value} \langle \mathit{true},c_1\rangle \qquad \langle s_1,\sigma,c_1\rangle \Downarrow_{value} \langle \sigma_1,c_2\rangle}{\langle \texttt{if}\ e\ \texttt{then}\ s_1\ \texttt{else}\ s_2\ \texttt{fi},\sigma,c\rangle \Downarrow_{value} \langle \sigma_1,c_2\rangle}$$

$$\text{IF-FALSE}\quad \frac{\langle e,\sigma,c\rangle \Downarrow_{value} \langle \mathit{false},c_1\rangle \qquad \langle s_2,\sigma,c_1\rangle \Downarrow_{value} \langle \sigma_2,c_2\rangle}{\langle \texttt{if}\ e\ \texttt{then}\ s_1\ \texttt{else}\ s_2\ \texttt{fi},\sigma,c\rangle \Downarrow_{value} \langle \sigma_2,c_2\rangle}$$

$$\text{SEQ}\quad \frac{\langle s_1,\sigma,c\rangle \Downarrow_{value} \langle \sigma_1,c_1\rangle \qquad \langle s_2,\sigma_1,c_1\rangle \Downarrow_{value} \langle \sigma_2,c_2\rangle}{\langle s_1\ ;\ s_2,\sigma,c\rangle \Downarrow_{value} \langle \sigma_2,c_2\rangle} \qquad \text{SKIP}\quad \frac{}{\langle \texttt{skip},\sigma,c\rangle \Downarrow_{value} \langle \sigma,c\rangle}$$

$$\text{WHILE-FALSE}\quad \frac{\langle e,\sigma,c\rangle \Downarrow_{value} \langle \mathit{false},c'\rangle}{\langle \texttt{while}\ e\ \texttt{do}\ s\ \texttt{od},\sigma,c\rangle \Downarrow_{value} \langle \sigma,c'\rangle}$$

$$\text{WHILE-TRUE}\quad \frac{\langle e,\sigma,c\rangle \Downarrow_{value} \langle \mathit{true},c_1\rangle \qquad \langle s_1,\sigma,c_1\rangle \Downarrow_{value} \langle \sigma_1,c_2\rangle \qquad \langle \texttt{while}\ e\ \texttt{do}\ s\ \texttt{od},\sigma_1,c_2\rangle \Downarrow_{value} \langle \sigma_2,c_3\rangle}{\langle \texttt{while}\ e\ \texttt{do}\ s\texttt{od},\sigma,c\rangle \Downarrow_{value} \langle \sigma_2,c_3\rangle}$$

**Figure 7: Augmented semantics of $\mathcal{L}$ for computing test-equivalence classes w.r.t. $\overset{t}{\sim}_{value}$. $v$ — variables, $n$ — integer values, $e$ — expressions, $s$ — statements, $\sigma$ — program states, $c$ — sets of expressions, *Modified* — predicate over expressions.**

occurrence of $l_1$. An example of applying this relation is given in Section 2.3.

The relation is formally defined through an augmented semantics of $\mathcal{L}$. As in Section 3.3, we chose this representation since it simultaneously defines an algorithm of identifying test-equivalence classes. The implementation of this semantics via program instrumentation is discussed in Section 5.

The semantics in Figure 8 extends the semantics in Figure 6 by defining the function $\Downarrow_{deps}$. It is parameterized by a predicate $\mathit{Inserted} : \mathit{Stmt} \to \mathbb{B}$ that marks the inserted assignment, a predicate $\mathit{Left} : \mathcal{V} \to \mathbb{B}$ that marks the left-hand side variable of the inserted assignment, and a predicate $\mathit{Right} : \mathcal{V} \to \mathbb{B}$ that marks the variables used in the right-hand side of the inserted assignment. $\Downarrow_{deps}$ additionally maintains (1) a set of locations $l$ representing test-equivalent insertions, (2) a set of locations $c$ that are executed after the last read/write of the variables involved in the inserted assignment, and (3) a boolean value $x$ that indicates if the inserted assignment was evaluated after the last read/write of the variables involved in the inserted assignment.

The augmented semantics describes an analysis algorithm that, for a given set of locations, "filters out" those that do not correspond to the test-equivalent insertions of a given assignment. For a program with inserted assignment $v := e$, the semantics in Figure 8 computes sequences of executed locations (stored in the set $c$) such that each sequence contains the inserted statement $v := e$ (the rule ASSIGN-INS) and all the rest of the statements in this sequence do not read/overwrite the variable $v$ and do not overwrite the variables in $e$ (the rule ASSIGN-NLR). When such a sequence is found, the set of locations executed in this sequence is intersected with the current set of test-equivalent insertions ($l \cap c$ in the rules VAR-LEFT-EXE and ASSIGN-LR-EXE). When the inserted assignment $v := e$ is not executed in such a sequence, the set of locations executed in this sequence is removed from the set of test-equivalent insertions ($l \setminus c$ in the rules VAR-LEFT-NEXE and ASSIGN-LR-NEXE), since for these locations this is no "matching" occurrence of $v := e$.

*Definition 3.5 (Dependency-based test-equivalence relation).* Let $\mathcal{P}$ be a set of programs, $t = (\sigma_{in}, \phi)$ be a test. $\overset{t}{\sim}_{deps} \subset \mathcal{P} \times \mathcal{P}$ is a *dependency-based test-equivalence relation* iff $p_1 \overset{t}{\sim}_{deps} p_2$ for $p_1, p_2 \in \mathcal{P}$ if there is program $p$ with locations $l_1, l_2$ such that $p_1 = p[{}^{l_1}/_{v := e; l_1}]$ and $p_2 = p[{}^{l_2}/_{v := e; l_2}]$ and $\langle p_1, \sigma_{in}, \{l_1, l_2\}, \emptyset, \mathit{false}\rangle \Downarrow_{value} \langle _, \{l_1, l_2\}, _, _\rangle$ given that $\mathit{Inserted} := \lambda s.\ s = \text{“}v := e\text{”}$, $\mathit{Left} := \lambda v'.\ v' = v$, $\mathit{Right} := \lambda v'.\ v' \in \mathit{Var}(e)$.

In this definition, we call two programs that differ in locations of an assignment insertion to be test-equivalent if the difference does not affect variable dependencies according to the semantics in Figure 6. Specifically, by passing the program $p_1$, the test input $\sigma_{in}$ and the set of locations $\{l_1, l_2\}$

$$\text{VAR-LEFT-EXE}\quad \frac{Left(v) \wedge x}{\langle v, \sigma, l, c, x\rangle \Downarrow_{deps} \langle \sigma(v), l \cap c, \emptyset, false\rangle}$$

$$\text{VAR-LEFT-NEXE}\quad \frac{Left(v) \wedge \neg x}{\langle v, \sigma, l, c, x\rangle \Downarrow_{deps} \langle \sigma(v), l \setminus c, \emptyset, x\rangle}$$

$$\text{VAR-NLEFT}\quad \frac{\neg Left(v)}{\langle v, \sigma, l, c, x\rangle \Downarrow_{deps} \langle \sigma(v), l, c, x\rangle}$$

$$\text{OP}\quad \frac{\langle e_1, \sigma, l, c, x\rangle \Downarrow_{deps} \langle n_1, l_1, c_1, x_1\rangle \qquad \langle e_2, \sigma, l_1, c_1, x_1\rangle \Downarrow_{deps} \langle n_1, l_1, c_1, x_1\rangle \qquad n_3 = n_1 \; op \; n_2}{\langle e_1 \; op \; e_2, \sigma, l, c, x\rangle \Downarrow_{deps} \langle n_3, l_2, c_2, x_2\rangle}$$

$$\text{ASSIGN-INS}\quad \frac{Inserted(v := e) \qquad \langle v := e, \sigma\rangle \Downarrow \sigma'}{\langle v := e, \sigma, l, c, x\rangle \Downarrow_{deps} \langle \sigma', l, c, true\rangle}$$

$$\text{ASSIGN-LR-EXE}\quad \frac{\neg Inserted(v := e) \wedge (Left(v) \vee Right(v)) \wedge x \qquad \langle v := e, \sigma\rangle \Downarrow \sigma'}{\langle v := e, \sigma, l, c, x\rangle \Downarrow_{deps} \langle \sigma', l \cap c, \emptyset, false\rangle}$$

$$\text{ASSIGN-LR-NEXE}\quad \frac{\neg Inserted(v := e) \wedge (Left(v) \vee Right(v)) \wedge \neg x \qquad \langle v := e, \sigma\rangle \Downarrow \sigma'}{\langle v := e, \sigma, l, c, x\rangle \Downarrow_{deps} \langle \sigma', l \setminus c, \emptyset, false\rangle}$$

$$\text{ASSIGN-NLR}\quad \frac{\neg Inserted(v := e) \wedge \neg Left(v) \wedge \neg Right(v) \qquad \langle e, \sigma, l, c, x\rangle \Downarrow_{deps} \langle n, l_1, c_1, x_1\rangle}{\langle v := e, \sigma, l, c, x\rangle \Downarrow_{deps} \langle \sigma[v \mapsto n], l_1, c_1, x_1\rangle}$$

$$\text{SEQ}\quad \frac{\langle s_1, \sigma, l, c, x\rangle \Downarrow_{deps} \langle \sigma_1, l_1, c_1, x_1\rangle \qquad \langle s_2, \sigma_1, l_1, c_1, x_1\rangle \Downarrow_{deps} \langle \sigma_2, l_2, c_2, x_2\rangle}{\langle s_1 \; ; \; s_2, \sigma, l, c, x\rangle \Downarrow_{deps} \langle \sigma_2, l_2, c_2, x_2\rangle}$$

$$\text{SKIP}\quad \frac{}{\langle \mathtt{skip}, \sigma, l, c, x\rangle \Downarrow_{deps} \langle \sigma, l, c, x\rangle}$$

$$\text{IF-TRUE}\quad \frac{\langle e, \sigma, l, c, x\rangle \Downarrow_{deps} \langle true, l_1, c_1, x_1\rangle \qquad \langle s_1, \sigma, l_1, c_1, x_1\rangle \Downarrow_{deps} \langle \sigma_1, l_2, c_2, x_2\rangle}{\langle \mathtt{if}\; e \;\mathtt{then}\; s_1 \;\mathtt{else}\; s_2 \;\mathtt{fi}, \sigma, c\rangle \Downarrow_{deps} \langle \sigma_1, l_2, c_2, x_2\rangle}$$

$$\text{IF-FALSE}\quad \frac{\langle e, \sigma, l, c, x\rangle \Downarrow_{deps} \langle false, l_1, c_1, x_1\rangle \qquad \langle s_2, \sigma, l_1, c_1, x_1\rangle \Downarrow_{deps} \langle \sigma_2, l_2, c_2, x_2\rangle}{\langle \mathtt{if}\; e \;\mathtt{then}\; s_1 \;\mathtt{else}\; s_2 \;\mathtt{fi}, \sigma, l, c, x\rangle \Downarrow_{deps} \langle \sigma_2, l_2, c_2, x_2\rangle}$$

$$\text{WHILE-TRUE}\quad \frac{\langle e, \sigma, l, c, x\rangle \Downarrow_{deps} \langle true, l_1, c_1, x_1\rangle \qquad \langle s_1, \sigma, l_1, c_1, x_1\rangle \Downarrow_{deps} \langle \sigma_1, l_2, c_2, x_2\rangle \qquad \langle \mathtt{while}\; e \;\mathtt{do}\; s \;\mathtt{od}, \sigma_1, l_2, c_2, x_2\rangle \Downarrow_{deps} \langle \sigma_2, l_3, c_3, x_3\rangle}{\langle \mathtt{while}\; e \;\mathtt{do}\; s \;\mathtt{od}, \sigma, l, c, x\rangle \Downarrow_{deps} \langle \sigma_2, l_3, c_3, x_3\rangle}$$

$$\text{WHILE-FALSE}\quad \frac{\langle e, \sigma, l, c, x\rangle \Downarrow_{deps} \langle false, l', c', x'\rangle}{\langle \mathtt{while}\; e \;\mathtt{do}\; s \;\mathtt{od}, \sigma, l, c, x\rangle \Downarrow_{deps} \langle \sigma, l', c', x'\rangle}$$

$$\text{NUM}\quad \frac{\langle n, \sigma\rangle \Downarrow n}{\langle n, \sigma, l, c, x\rangle \Downarrow_{deps} \langle n, l, c, x\rangle}$$

**Figure 8: Augmented semantics of $\mathcal{L}$ for computing test-equivalence classes w.r.t. $\overset{t}{\sim}_{deps}$. $v$ — variables, $n$ — integer values, $e$ — expressions, $s$ — statements, $\sigma$ — program states, $l, c$ — sets of locations, $x$ — boolean values, *Inserted* — predicate over statements, *Left*, *Right* — predicates over variables.**

as the arguments to $\Downarrow_{deps}$, we obtain $\{l_1, l_2\}$ as the result.

PROPOSITION 3.6. *The relation $\overset{t}{\sim}_{deps}$ is a test-equivalence relation according to Definition 1.1.*

The proposition above formally states that (1) $\overset{t}{\sim}_{deps}$ is an equivalence relations and (2) if two programs that differ only in locations in which the same assignment statement is inserted are such that these differences do not impact dynamic variable dependencies (according to the semantics in Figure 8), then these two programs either both pass the test of both fail the test.

## 3.5 Composing relations

Several test-equivalence relations can be composed in a mutually-reinforcing fashion. By combining several relations, we can produce a more effective (coarse-grained) partitioning of program modifications into test-equivalence classes.

*Definition 3.7 (Composition of relations).* Let $\mathcal{P}$ be a search space and $\overset{t}{\sim}_1, \overset{t}{\sim}_2, ..., \overset{t}{\sim}_n$ be test-equivalence relations in $\mathcal{P}$. A *composition* of $\overset{t}{\sim}_1, \overset{t}{\sim}_2, ..., \overset{t}{\sim}_n$ is a test-equivalence relation $\overset{t}{\sim}_*$ that is the transitive closure of the union of $\overset{t}{\sim}_1, \overset{t}{\sim}_2, ..., \overset{t}{\sim}_n$:

$$\overset{t}{\sim}_* := (\bigcup_i \overset{t}{\sim}_i)^*$$

$$\text{VAR-LEFT-EXE}\quad \frac{\mathit{Left}(v) \land x}{\langle v,\sigma,l,c,x\rangle \Downarrow_* \langle \sigma(v), l \cap c, \emptyset, \mathit{false}\rangle}$$

$$\text{VAR-LEFT-NEXE}\quad \frac{\mathit{Left}(v) \land \neg x}{\langle v,\sigma,l,c,x\rangle \Downarrow_* \langle \sigma(v), l \setminus c, \emptyset, x\rangle}$$

$$\text{VAR-NLEFT}\quad \frac{\neg \mathit{Left}(v)}{\langle v,\sigma,l,c,x\rangle \Downarrow_* \langle \sigma(v), l, c, x\rangle}$$

$$\text{OP}\quad \frac{\langle e_1,\sigma,l,c,x\rangle \Downarrow_* \langle n_1,l_1,c_1,x_1\rangle \qquad \langle e_2,\sigma,l_1,c_1,x_1\rangle \Downarrow_* \langle n_1,l_1,c_1,x_1\rangle \qquad n_3 = n_1 \mathbin{op} n_2}{\langle e_1 \mathbin{op} e_2,\sigma,l,c,x\rangle \Downarrow_* \langle n_3,l_2,c_2,x_2\rangle}$$

$$\text{ASSIGN-INS}\quad \frac{\mathit{Inserted}(v := e) \qquad \langle v := e,\sigma\rangle \Downarrow \sigma'}{\langle v := e,\sigma,l,c,x\rangle \Downarrow_* \langle \sigma',l,c,\mathit{true}\rangle}$$

$$\text{ASSIGN-LR-EXE}\quad \frac{\neg \mathit{Inserted}(v := e) \land (\mathit{Left}(v) \lor \mathit{Right}(v)) \land x \qquad \langle v := e,\sigma\rangle \Downarrow \sigma'}{\langle v := e,\sigma,l,c,x\rangle \Downarrow_* \langle \sigma', l \cap c, \emptyset, \mathit{false}\rangle}$$

$$\text{ASSIGN-LR-NEXE}\quad \frac{\neg \mathit{Inserted}(v := e) \land (\mathit{Left}(v) \lor \mathit{Right}(v)) \land \neg x \qquad \langle v := e,\sigma\rangle \Downarrow \sigma'}{\langle v := e,\sigma,l,c,x\rangle \Downarrow_* \langle \sigma', l \setminus c, \emptyset, \mathit{false}\rangle}$$

$$\text{ASSIGN-NLR}\quad \frac{\neg \mathit{Inserted}(v := e) \land \neg \mathit{Left}(v) \land \neg \mathit{Right}(v) \qquad \langle e,\sigma,l,c,x\rangle \Downarrow_* \langle n,l_1,c_1,x_1\rangle}{\langle v := e,\sigma,l,c,x\rangle \Downarrow_* \langle \sigma[v \mapsto n],l_1,c_1,x_1\rangle}$$

$$\text{SEQ}\quad \frac{\langle s_1,\sigma,l,c,x\rangle \Downarrow_{value*} \langle \sigma_1,l_1,c_1,x_1\rangle \qquad \langle s_2,\sigma_1,l_1,c_1,x_1\rangle \Downarrow_{value*} \langle \sigma_2,l_2,c_2,x_2\rangle}{\langle s_1 \,;\, s_2,\sigma,l,c,x\rangle \Downarrow_* \langle \sigma_2,l_2,c_2,x_2\rangle}$$

$$\text{SKIP}\quad \frac{}{\langle \mathtt{skip},\sigma,l,c,x\rangle \Downarrow_* \langle \sigma,l,c,x\rangle}$$

$$\text{IF-TRUE}\quad \frac{\langle e,\sigma,l,c,x\rangle \Downarrow_* \langle \mathit{true},l_1,c_1,x_1\rangle \qquad \langle s_1,\sigma,l_1,c_1,x_1\rangle \Downarrow_{value*} \langle \sigma_1,l_2,c_2,x_2\rangle}{\langle \mathtt{if}\ e\ \mathtt{then}\ s_1\ \mathtt{else}\ s_2\ \mathtt{fi},\sigma,c\rangle \Downarrow_* \langle \sigma_1,l_2,c_2,x_2\rangle}$$

$$\text{IF-FALSE}\quad \frac{\langle e,\sigma,l,c,x\rangle \Downarrow_* \langle \mathit{false},l_1,c_1,x_1\rangle \qquad \langle s_2,\sigma,l_1,c_1,x_1\rangle \Downarrow_{value*} \langle \sigma_2,l_2,c_2,x_2\rangle}{\langle \mathtt{if}\ e\ \mathtt{then}\ s_1\ \mathtt{else}\ s_2\ \mathtt{fi},\sigma,l,c,x\rangle \Downarrow_* \langle \sigma_2,l_2,c_2,x_2\rangle}$$

$$\text{WHILE-TRUE}\quad \frac{\langle e,\sigma,l,c,x\rangle \Downarrow_* \langle \mathit{true},l_1,c_1,x_1\rangle \qquad \langle s_1,\sigma,l_1,c_1,x_1\rangle \Downarrow_{value*} \langle \sigma_1,l_2,c_2,x_2\rangle \qquad \langle \mathtt{while}\ e\ \mathtt{do}\ s\ \mathtt{od},\sigma_1,l_2,c_2,x_2\rangle \Downarrow_{value*} \langle \sigma_2,l_3,c_3,x_3\rangle}{\langle \mathtt{while}\ e\ \mathtt{do}\ s\ \mathtt{od},\sigma,l,c,x\rangle \Downarrow_* \langle \sigma_2,l_3,c_3,x_3\rangle}$$

$$\text{WHILE-FALSE}\quad \frac{\langle e,\sigma,l,c,x\rangle \Downarrow_* \langle \mathit{false},l',c',x'\rangle}{\langle \mathtt{while}\ e\ \mathtt{do}\ s\ \mathtt{od},\sigma,l,c,x\rangle \Downarrow_* \langle \sigma,l',c',x'\rangle}$$

$$\text{NUM}\quad \frac{\langle n,\sigma\rangle \Downarrow n}{\langle n,\sigma,l,c,x\rangle \Downarrow_* \langle n,l,c,x\rangle}$$

$$\text{STMT-L}\quad \frac{\langle s,\sigma,l,c,x\rangle \Downarrow_* \langle \sigma',l',c',x'\rangle \qquad c'' = \lambda x.\ \text{if } x = s \text{ then } c'(x) \cap \Pi^{value}_{\sigma,n}(\varepsilon) \text{ else } c'(x)}{\langle s,\sigma,l,c,x\rangle \Downarrow_{value*} \langle \sigma',l',c'',x'\rangle}$$

**Figure 9: Augmented semantics of $\mathcal{L}$ for computing test-equivalence classes w.r.t. $\overset{t}{\sim}^*$. $v$ — variables, $n$ — integer values, $e$ — expressions, $s$ — statements, $\sigma$ — program states, $l, c$ — functions from locations to sets of expressions, $x$ — boolean values, *Inserted* — predicate over statements, *Left*, *Right* — predicates over variables, $\varepsilon$ — right-hand side of inserted assignment.**

In this work, we define a test-equivalence relation $\overset{t}{\sim}^*$ as a composition of the relations $\overset{t}{\sim}_{value}$ and $\overset{t}{\sim}_{deps}$:

$$\overset{t}{\sim}^* := (\overset{t}{\sim}_{value} \cup \overset{t}{\sim}_{deps})^*$$

An example of applying $\overset{t}{\sim}^*$ is given in Section 2.3. It is straightforward to define an augmented semantics of $\mathcal{L}$ for $\overset{t}{\sim}^*$ by combining the semantics in Figure 7 and Figure 8 as shown in Figure 9. For a given program $p$ with an assignment $v := \varepsilon$, it identifies a mapping $\{\, l_1 \mapsto \mathcal{E}_1, \ldots, l_k \mapsto \mathcal{E}_k \,\}$ that denotes all test-equivalent insertions of assignments of $v := e'$ at each location $l_i$ so that $e' \in \mathcal{E}_i$. In this semantics, $l$ and $c$ denote such mappings, $l \cap c$ denotes $\lambda x.\ l(x) \cup c(x)$ and $l \setminus c$ denotes $\lambda x.\ l(x) \setminus c(x)$.

$M(s_1; s_2) = \cup_{s' \in M(s_1)}\{\ s'; s_2\ \} \ \cup\ \cup_{s' \in M(s_2)}\{\ s_1; s'\ \}$

$M(\texttt{if}\ e\ \texttt{then}\ s_1 \texttt{else}\ s_2\ \texttt{fi}) = \cup_{s' \in M(s_1)}\{\ \texttt{if}\ e\ \texttt{then}\ s' \texttt{else}\ s_2\ \texttt{fi}\ \} \cup \cup_{s' \in M(s_2)}\{\ \texttt{if}\ e\ \texttt{then}\ s_1 \texttt{else}\ s'\ \texttt{fi}\ \}$

$M(\texttt{while}\ e\ \texttt{do}\ s\ \texttt{od}) = \cup_{s' \in M(s)}\{\ \texttt{while}\ e\ \texttt{do}\ s'\ \texttt{od}\}$

$M(s \mid s \text{ is not a sequence}) = M_{EXPRESSION}(s) \cup M_{REFINEMENT}(s) \cup M_{GUARD}(s) \cup M_{ASSIGNMENT}(s)$

$M_{EXPRESSION}(v := e) = \cup_{e' \in \mathcal{E}}\{\ v := e'\ \}$

$M_{EXPRESSION}(\texttt{if}\ e\ \texttt{then}\ s_1\ \texttt{else}\ s_2\ \texttt{fi}) = \cup_{e' \in \mathcal{E}}\{\ \texttt{if}\ e'\ \texttt{and}\ e'\ \texttt{then}\ s_1\ \texttt{else}\ s_2\ \texttt{fi}\ \}$

$M_{EXPRESSION}(\texttt{while}\ e\ \texttt{do}\ s\ \texttt{od}) = \cup_{e' \in \mathcal{E}}\{\ \texttt{while}\ e'\ \texttt{do}\ s\ \texttt{od}\}$

$M_{REFINEMENT}(\texttt{if}\ e\ \texttt{then}\ s_1\ \texttt{else}\ s_2\ \texttt{fi}) = \cup_{e' \in \mathcal{E}}\{\ \texttt{if}\ e\ \texttt{and}\ e'\ \texttt{then}\ s_1\ \texttt{else}\ s_2\ \texttt{fi}, \texttt{if}\ e\ \texttt{or}\ e'\ \texttt{then}\ s_1\ \texttt{else}\ s_2\ \texttt{fi}\ \}$

$M_{REFINEMENT}(\texttt{while}\ e\ \texttt{do}\ s\ \texttt{od}) = \cup_{e' \in \mathcal{E}}\{\ \texttt{while}\ e\ \texttt{and}\ e'\ \texttt{do}\ s\ \texttt{od}, \texttt{while}\ e\ \texttt{or}\ e'\ \texttt{do}\ s\ \texttt{od}\}$

$M_{GUARD}(v := e) = \cup_{e' \in \mathcal{E}}\{\ \texttt{if}\ e'\ \texttt{then}\ v := e\ \texttt{else skip fi}\ \}$

$M_{ASSIGNMENT}(s) = \cup_{e' \in \mathcal{E}, v' \in \mathcal{V}}\{\ v' := e';\ s\ \}$

**Figure 10: Search space definition via transformation schemas $M$.**

# 4 PATCH GENERATION

Automated program repair techniques search for patches in spaces of candidate program modifications. A search space in program repair is defined as in the following.

*Definition 4.1 (Search space).* A *search space* is a finite set of syntactically different programs obtained by applying a given transformation function $M : \mathcal{L} \to 2^{\mathcal{L}}$ to the buggy program.

Previous systems (e.g. SPR/Prophet [Long and Rinard 2015, 2016b] and SemFix/Angelix [Mechtaev et al. 2016; Nguyen et al. 2013]) defined their search spaces through *parameterized transformation schemas* such that each schema transforms a given program into a program with "holes" and the "holes" are filled with expressions using a program synthesizer. We define our search space in a similar fashion via the function $M$ in Figure 10 ($\mathcal{E}$ indicates the synthesized expressions).

*Definition 4.2 (Optimal program repair).* Let $T$ be a test-suite (a set of tests), $p \in \mathcal{L}$ be a buggy program ($\exists t \in T.\ \neg Pass[p, t]$), $M : \mathcal{L} \to 2^{\mathcal{L}}$ be a transformation function, $\mathcal{P} := M(p)$ be the corresponding search space, $\kappa : \mathcal{P} \to \mathbb{R}$ be a cost function. The goal of *optimal program repair* is to find a repair $p' \in \mathcal{P}$ such that $\forall t \in T.\ Pass[p', t]$ and $\kappa(p')$ is minimal among all such programs.

Our patch generation algorithm systematically explore the search space by (1) evaluating candidates in the order defined by the prioritization (cost function) starting from the highest priority patch and (2) skipping redundant executions by on-the-fly identifying test-equivalence classes w.r.t. a given test-equivalence relation.

**ALGORITHM 2:** Systematic exploration with partitioning into test-equivalence classes

**Input**: search space $\mathcal{P}$, cost function $\kappa$, test-suite $T$, test-equivalence relation $\overset{t}{\sim}$
**Output**: ordered set of repairs $R$

$R := \emptyset$;
**foreach** $t \in T$ **do**
  $C(t), \overline{C}(t) := \emptyset, \emptyset$;
**while** $\mathcal{P} \neq \emptyset$ **do**
  p := $pick(\mathcal{P}, \kappa)$;
  **if** $\exists t.\ \bigvee_{c \in \overline{C}(t)} p \in c$ **then**
    $\mathcal{P} := \mathcal{P} \setminus \{p\}$;
    **continue**;
  **foreach** $t \in T$ **do**
    **if** $\bigvee_{c \in C(t)} p \in c$ **then**
      **continue**;
    $isPassing$, $[p] := \widetilde{eval}(p, t, \mathcal{P}, \overset{t}{\sim})$;
    **if** $isPassing$ **then**
      $C(t) := C(t) \cup \{[p]\}$;
    **else**
      $\overline{C}(t) := \overline{C}(t) \cup \{[p]\}$;
      **break**;
  **if** $\forall t.\ \bigvee_{c \in C(t)} p \in c$ **then**
    $R := R \cup \{p\}$;
  $\mathcal{P} := \mathcal{P} \setminus \{p\}$;

In order to abstract over various optimal synthesis methodologies, we assume that there is a function *pick* that for a given set of programs $\mathcal{P}$ and a cost function $\kappa$, returns a program from $\mathcal{P}$ with the minimal value of $\kappa$.

The overall workflow of our approach is described in Algorithm 2. Our algorithm takes a patch space (Definition 4.1), a cost function, a test-suite and a test-equivalence relation as inputs and outputs a sequence of search space elements that pass all the given tests ordered according to the cost function. The algorithm maintains sets $C$ and $\overline{C}$ for each test. $C(t)$ is a set of test-equivalence classes (therefore, $C$ is a set of sets of programs) in which all candidates pass $t$; $\overline{C}(t)$ is the corresponding set of failing test-equivalence classes.

First, our algorithm initializes the list of output repairs $R$ and the passing and failing test-equivalence classes $C$ and $\overline{C}$ for all tests $t$. Second, it iterates through the search space by (1) picking the best (the highest cost according to $\kappa$) remaining candidate using *pick* and (2) evaluating the candidate with the tests and computing test-equivalence classes.

For a given candidate, in order to identify the result of a test execution and the corresponding test-equivalence class, the algorithm evaluates the candidate using the function $\widetilde{eval}$. The function $\widetilde{eval}$ takes a program $p$, a test $t$, a search space $\mathcal{P}$ and a test-equivalence relation $\overset{t}{\sim}$ and returns the result of executing $p$ with $t$ (as a boolean value *isPassing*) and a set of programs $[p]$ such that (1) $p \in [p]$, (2) $[p] \subset \mathcal{P}$ and (3) $[p]$ is a test-equivalence class of $\mathcal{P}$ w.r.t. the relation $\overset{t}{\sim}$. The concrete implementation of $\widetilde{eval}$ depends on the relation $\overset{t}{\sim}$ and is formally described for the relations $\overset{t}{\sim}_{value}$ and $\overset{t}{\sim}_{deps}$ in Definition 4.3 and Definition 4.4 respectively.

The test-equivalence classes are used at two steps of search space exploration. First, after a next candidate is picked, the algorithm checks if the candidate belongs to any of the existing failing classes (line 6). If the candidate is in a failing class of at least one test, evaluation of this candidate is omitted. Second, after a next test is selected for evaluating a candidate, the algorithm checks if the candidate is in a passing class of the given test (line 10). If a candidate is in a passing class of a test, then the algorithm omits execution of this candidate with this test. We now discuss the function $\widetilde{eval}$ specifically for the two notions of test-equivalence we have studied, the value-based test-equivalence, and dependence based test-equivalence

*Definition 4.3 (Value-based test-equivalence analysis).* Let $\mathcal{P}$ be a search space, $p \in \mathcal{P}$ be a program, $t = (\sigma_{in}, \phi)$ be a test. Let $e$ be an expression in $p$ such that $\exists p' \in \mathcal{P} \exists e' \in Expr.\, p' = p[^e/_{e'}]$. Then, value-based test-equivalence analysis $\widetilde{eval}$ is defined as follows:

$$\widetilde{eval}(p, t, \mathcal{P}, \overset{t}{\sim}_{value}) = (\phi(\sigma_{out}), \bigcup_{e' \in C} \{p[^e/_{e'}]\}), \text{ given that}$$
$$Modified := \lambda x.\, x = e,$$
$$\mathcal{E} := \{e' \mid \exists p' \in \mathcal{P}.\, p' = p[^e/_{e'}]\},$$
$$\langle p, \sigma_{in}, \mathcal{E}\rangle \Downarrow_{value} \langle \sigma_{out}, C\rangle$$

In this analysis, we identify a test-equivalence class of a program with an expression $e$ in the space of all programs in $\mathcal{P}$ that differ only in $e$. The test-equivalence class is computed by passing the set of all "alternative" expressions $\mathcal{E}$ as an argument of $\Downarrow_{value}$. Note that in this definition we explicitly select an expression $e$ substitution of which are analyzed for test-equivalence. For each element of a search space produced by the transformation function $M$ in Figure 10, there is always at most one such $e$. We now discuss the function $\widetilde{eval}$ for dependency-based test-equivalence.

*Definition 4.4 (Dependency-based test-equivalence analysis).* Let $\mathcal{P}$ be a search space, $p \in \mathcal{P}$ be a program, $t = (\sigma_{in}, \phi)$ be a test. Let $p'$ be a program, $l_1$ be a location such that $p = p'[^{l_1}/_{\mathsf{v} := \mathsf{e}; l_1}]$ and $\exists p'' \in \mathcal{P}.\, \exists l_2 \in p''.\, p'' = p'[^{l_2}/_{\mathsf{v} := \mathsf{e}; l_2}]$. Then, dependency-based test-equivalence analysis $\widetilde{eval}$ is:

$$\widetilde{eval}(p, t, \mathcal{P}, \overset{t}{\sim}_{deps}) = (\phi(\sigma_{out}), \bigcup_{l \in L'} \{p'[^l/_{\mathsf{v} := \mathsf{e}; l}]\}), \text{ given that}$$
$$Inserted := \lambda s.\, s = \text{“}v := e\text{”},$$
$$Left := \lambda v'.\, v' = v,$$
$$Right := \lambda v'.\, v' \in Var(e),$$
$$L := \{\, l \mid l \in p' \wedge \exists p'' \in \mathcal{P}.\, p'' = p'[^l/_{\mathsf{v} := \mathsf{e}; l}]$$
$$\langle p, \sigma_{in}, L, \emptyset, false\rangle \Downarrow_{value} \langle \sigma_{out}, L', _, _\rangle$$

In this analysis, we identify a test-equivalence class of a program with an assignment $v := e$ in the space of all programs in $\mathcal{P}$ that differ only in locations of this assignment. The test-equivalence class is computed by passing the set of all "alternative" locations $L$ as an argument of $\Downarrow_{deps}$.

Finally, note that Algorithm 2 can be used in different ways. The output of the algorithm is a sequence of plausible patches $R$ ordered according to the function $\kappa$. A sequence of repairs can used to provide several patch suggestions for developers. The number of suggested repairs can be controlled by introducing a limit and breaking from the main loop when the required number of plausible patches is found. Certain applications may require generation of all plausible patches (e.g. in order to narrow candidates through test generation [Shriver et al. 2017]). In this case, the algorithm can be modified so that it outputs whole test-passing partitions instead of single patches.

## 5 IMPLEMENTATION

We have implemented the described approach in an tool called f1x (pronounced as [ɛf-wʌn-ɛks]) for the C programming language.

*Analysis.* Our implementation of the two test-equivalence analyses is built upon a combination of static (source code) and dynamic instrumentation. Specifically, to implement the augmented semantics in Section 3.3 for the relation $\overset{t}{\sim}_{value}$, we apply the transformation schemas $M$ (Figure 10) to the source code of the buggy program and replace "holes" with calls to a procedure implementing the value-projection operator. To implement the augmented semantics in Section 3.4 for the relation $\overset{t}{\sim}_{deps}$, we implemented a dynamic instrumentation using Pin [Luk et al. 2005] that tracks reads and writes of the variables involved in assignment synthesis.

*Search space.* The goal of this work was to design and evaluate test-equivalence relations for transformations used in existing program repair systems. Our system combines the transformation schemas of SPR/Prophet and Angelix (we studied implementation of these systems in order to closely reproduce their search spaces), described as follows.

**EXPRESSION** Modify an existing side-effect free integer expression or condition (adopted from Angelix). A variant of $M_{EXPRESSION}$ in Figure 10 for C programs. Partitioned into test-equivalence classes based on the expression values using $\overset{t}{\sim}_{value}$.

**REFINEMENT** Append a disjunct/conjunct to an existing condition (adopted from Prophet). A variant of $M_{REFINEMENT}$ in Figure 10 for C programs. Partitioned into test-equivalence classes based on the condition values using $\overset{t}{\sim}_{value}$.

**GUARD** Add an if-guard for an existing statement (adopted from Angelix and Prophet). A variant of $M_{GUARD}$ in Figure 10 for C programs. Partitioned into test-equivalence classes based on the condition values using $\overset{t}{\sim}_{value}$.

**ASSIGNMENT** Insert an assignment statement (adopted from Prophet[7]). A variant of $M_{ASSIGNMENT}$ in Figure 10 for C programs. Partitioned into test-equivalence classes using $\overset{t}{\sim}_{*}$.

**INITIALIZATION** Insert memory initialization (adopted from Prophet). Not partitioned.

**FUNCTION** Replace a function call with another function (adopted from Prophet). Not partitioned.

The two last transformation schemas adopted from Prophet are not partitioned by our algorithm, since they generate relatively small search spaces. Our transformations differ from that of SPR/Prophet in the following ways: (1) Prophet implements a transformation schema for inserting guarded return statements. Although our algorithm can partition these transformations using the relation $\overset{t}{\sim}_{value}$, such transformations were shown to frequently generate overfitting patches [Tan et al. 2016] and therefore we exclude them from our search space; (2) Prophet implements a transformation that copies existing program statements. Since such statements can be arbitrarily complex and they cannot be partitioned by our algorithm, we do not include this transformation.

*Cost function.* Several techniques have been proposed to increase the probability of generating correct repairs by prioritizing patches [D'Antoni et al. 2016; Long and Rinard 2016b; Mechtaev et al. 2015]. For our system we implement an approach that assigns higher priority to smaller changes [Mechtaev et al. 2015]:

$$\kappa(p) := distance(p, p_{orig})$$

[7]Prophet generates new assignments by copying and modifying existing assignments. Instead, f1x *synthesizes* assignments and therefore its search space includes a superset of assignments that can be generated by Prophet.

where $p$ is a patched program (an element of the search space), $p_{orig}$ is the original program, *distance* is defined as the number of added, modified and deleted AST nodes.

## 6 EXPERIMENTAL EVALUATION

We evaluate our approach in terms of the following research questions:

**(RQ1)** What are the effectiveness and efficiency of our approach compared with state-of-the-art program repair systems?

**(RQ2)** Does our approach scale to larger search spaces compared with state-of-the-art systems? Does test-equivalence relation enable higher scalability of our implementation?

**(RQ3)** What is the impact of each test-equivalence relation on the number of test executions performed by our algorithm?

### 6.1 Evaluation setup

Our evaluation compares f1x against three repair approaches: Angelix, Prophet and GenProg-AE. These repair techniques are chosen as they use different repair algorithms including symbolic analysis (Angelix), machine-learning (Prophet) and genetic algorithm (GenProg). We evaluate all repair approaches on the GenProg ICSE'12 benchmark [Le Goues et al. 2012a] for our evaluation because it includes defects from large real-world projects, and was designed for systematic evaluation of program repair tools. Moreover, the test suites in this benchmark were independently augmented to prevent repair tools from generating implausible patches [Qi et al. 2015]. The benchmark consists of 105 defects from eight subjects (i.e. libtiff, lighttpd, PHP, gmp, gzip, python, wireshark, and fbc) which have developer-written test suites. Table 1 shows the statistics of each evaluated subject. The column "Execution cost" denotes the time taken to execute the test-suite for a given subject.

We selected the following systems and their configurations for evaluation:

**F1X** f1x that implements our test-equivalence partitioning technique.

**F1X$^E$** f1x$^E$ is a variant of f1x that enumerates changes without test-equivalence partitioning.

**ANG** Angelix 1.1 [Mechtaev et al. 2016] that implements a symbolic path exploration and prioritizes syntactically small changes.

**PR** Prophet 0.1 [Long and Rinard 2016b] that implements value search (a variant of path exploration) for conditional expressions and patch prioritization based on machine learning.

**PR*** Prophet* that is a variant of Prophet that disables transformations for (1) inserting overfitting return insertions and (2) copying complex statements except for assignments.

**GP** GenProg-AE 3.0 [Weimer et al. 2013] that implements a group of analysis techniques to avoid evaluating functionally-equivalent patches (as opposite to test-equivalent as in our approach).

The search space of F1X/F1X$^E$ is effectively the combination of the search spaces of PR* and ANG. We run two variants of our implementation: one variant with the introduced partitioning algorithm (F1X) and another variant without partitioning (F1X$^E$) to evaluate implementation-independent effect of the partitioning algorithm.

We run all the configurations (F1X, F1X$^E$, ANG, PR, PR*, GP) in two modes:

**Stop-after-first-found** The algorithm terminates after finding the first patch. This mode represents the usual program repair usage scenario.

**Full exploration** The algorithm terminates after searching through the entire search space. This mode allows us to obtain data that is independent on (1) the exploration order and (2) whether a plausible patch is present in the search space.

We reuse the configurations from previous studies for running Angelix, Prophet and GenProg-AE [Qi et al. 2015; Weimer et al. 2013]. As Prophet takes a correctness model as input to prioritizes patches akin to the provided model, we used the default model that is publicly available[8]. We conduct all experiments on Intel® Xeon™ CPU E5-2660 machines running Ubuntu 14.04, and use a 10 hours timeout for running each configuration.

[8]Prophet website: http://rhino.csail.mit.edu/prophet-rep/

**Table 1: Subject programs and their basic statistics**

| Program | Description | LOC | Defects | Tests | Execution cost (sec) |
|---|---|---|---|---|---|
| libtiff | Image processing library | 77K | 24 | 78 | 8.18 |
| lighttpd | Web server | 62K | 9 | 295 | 29.75 |
| php | Interpreter | 1046K | 44 | 8471 | 427.25 |
| gmp | Math library | 145K | 2 | 146 | 53.52 |
| gzip | Data compression utility | 491K | 5 | 12 | 0.43 |
| python | Interpreter | 407K | 11 | 35 | 156.46 |
| wireshark | Network packet analyzer | 2814K | 7 | 63 | 9.92 |
| fbc | Compiler | 97K | 3 | 773 | 240.27 |

## 6.2 Effectiveness and efficiency (RQ1)

Table 2 summarizes the effectiveness results for F1X, F1X$^E$, ANG, PR, PR* and GP executed in the stop-after-first-found mode. The second through seventh columns denote the number of plausible patches generated by each repair approach, while the eighth through thirteenth columns represent the number of patches syntactically equivalent to the human patches. As Angelix does not support lighttpd, python and fbc, the corresponding cells for these subjects are marked with "-". The overall results illustrate that F1X generates the highest number of plausible patches compared to all other evaluated repair approaches. The "Equivalent to human" column in table 2 shows that F1X generates 8 more human-like patches than F1X$^E$, 6 more human-like patches than ANG, 1 more human-like patch than PR, 2 more human-like patches than PR* and 13 more human-like patches than GP.

We attribute the high number of patches generated by F1X to the larger patch space supported by F1X compared to other approaches. Since F1X combines the search spaces of ANG and PR*, it fixes all defects that are fixed by either of these tools. Note that F1X finds more patches than F1X$^E$ within the time limit due to the performance gain from our partitioning.

Figure 11 illustrates the average patch generation time for the configurations. The x-axis of Figure 11 represents the eight subjects in the benchmark, while the y-axis shows the average time taken to generate a patch for all defects for a given subject where each bar depicts a patch generation approach. Overall, the average patch generation time for F1X is significantly shorter than all other repair approaches. For instance, F1X requires only 121 seconds on average to generate a patch for libtiff, while ANG takes 1262 seconds (F1X is $\frac{1262}{121}$=10.5X faster than ANG). Meanwhile, PR* takes 1701 seconds on average to produce a patch for libtiff (F1X is $\frac{1701}{121}$=14X faster than PR*). Notably, F1X is 16X faster than GP for libtiff (GP takes 1940 seconds on average to generate a patch for libtiff). The average patch generation time for PR is slightly higher compared to PR* as it searches through a slightly larger patch space.

The results shown in Figure 11 validate our claim that F1X is able to achieve significant improvement on the patch generation time due to its efficient search algorithm. F1X and F1X$^E$ demonstrate a comparable average time of patch generation because F1X$^E$ finds a subset of patches found by F1X that appears early in the sequence of explored candidates.

## 6.3 Exploration Speed (RQ2)

*Definition 6.1 (Explored candidates).* We say that a candidate patch is *explored* if the algorithm identified whether the patch passes all given tests or fails at least one. Note that we only consider candidate patches in which the source code modification is executed by all given failing tests.

Table 3 shows the exploration statistics for F1X, F1X$^E$, PR, PR* and GP (we exclude ANG because the search space for Angelix is encoded via logical constraints). The second through sixth columns depicts the data for the stop-after-first-found mode, while the seventh through eleventh columns represents the data for the full exploration mode. Each cell in the second through eleventh columns is of the form $\frac{X}{Y}$=Z where X represents the average number of "explored candidates" by a repair approach, Y represents the average number of test executions performed by a repair approach, Z denotes the exploration speed (computed by the ratio of the number of "explored candidates" over the number of test executions). The average for the stop-after-first-found is computed among the fixed defects, whereas the average for the full exploration mode is computed among all defects.

**Table 2: Effectiveness of program repair approaches.**

| Subject | Plausible | | | | | | Equivalent to human | | | | | |
|---|---|---|---|---|---|---|---|---|---|---|---|---|
| | F1X | F1X$^E$ | ANG | PR | PR* | GP | F1X | F1X$^E$ | ANG | PR | PR* | GP |
| libtiff | 13 | 10 | 10 | 5 | 3 | 5 | 5 | 3 | 3 | 2 | 1 | 0 |
| lighttpd | 5 | 3 | - | 4 | 4 | 4 | 0 | 0 | - | 0 | 0 | 0 |
| php | 15 | 7 | 10 | 18 | 18 | 7 | 6 | 3 | 4 | 10 | 10 | 2 |
| gmp | 2 | 1 | 2 | 2 | 2 | 1 | 2 | 1 | 2 | 1 | 1 | 0 |
| gzip | 3 | 2 | 2 | 2 | 2 | 2 | 2 | 0 | 1 | 1 | 1 | 0 |
| python | 5 | 1 | - | 6 | 6 | 3 | 0 | 0 | - | 0 | 0 | 1 |
| wireshark | 4 | 4 | 4 | 4 | 4 | 4 | 0 | 0 | 0 | 0 | 0 | 0 |
| fbc | 1 | 1 | - | 1 | 1 | 1 | 1 | 1 | - | 1 | 1 | 0 |
| Overall | 49 | 28 | 28 | 42 | 40 | 27 | 16 | 8 | 10 | 15 | 14 | 3 |

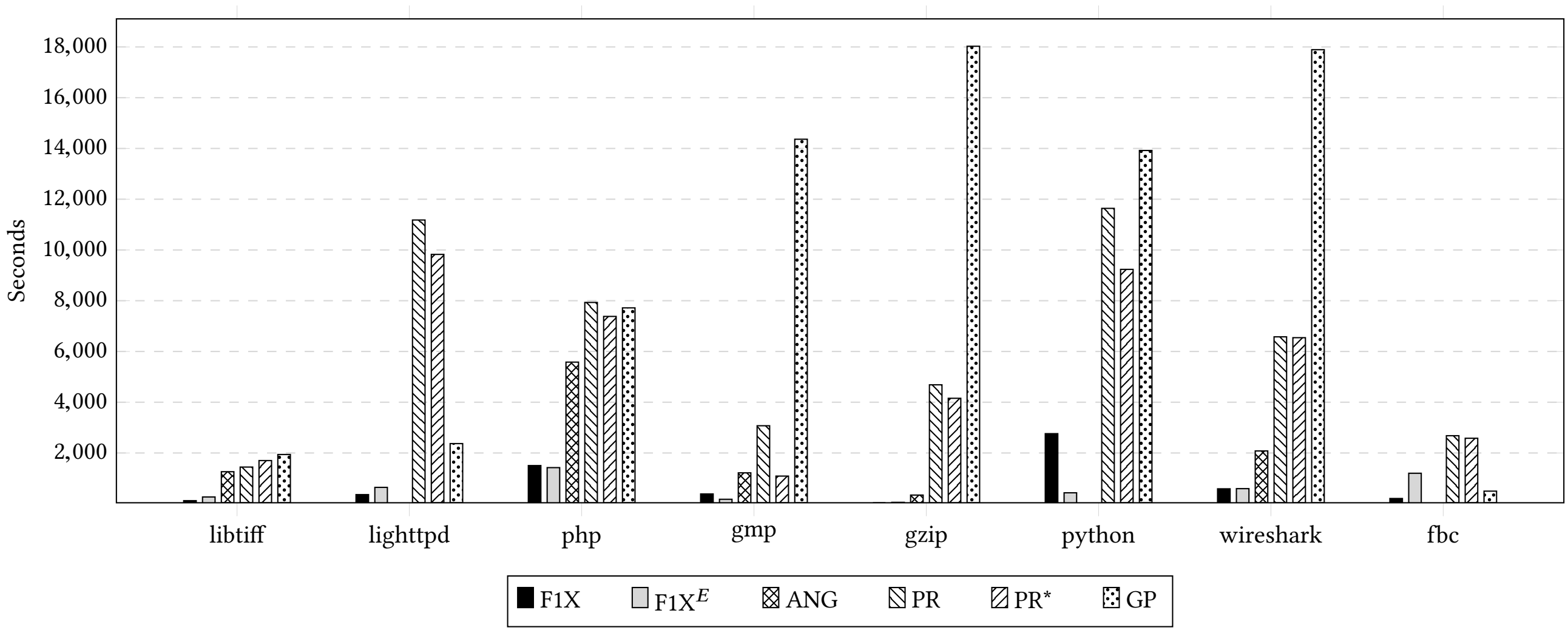

**Figure 11: Average patch generation time.**

In general, F1X has at least an order of magnitude higher exploration speed compared to all other patch generation approaches in both the stop-after-first-found mode and the full exploration mode. For example, in the full exploration mode, F1X requires on average only 620 test executions to explores 689925 candidates for wireshark. For the same subject, PR requires 13099 test executions for exploring 23043 candidates, PR* requires 17442 test executions for exploring 18554 candidates and GP requires 2213 test executions for exploring 2008 candidates.

To enable generation of more patches, an ideal repair approach should scale to larger spaces by exploring more candidates within the time budget. The data in Table 3 shows that F1X scales to larger search spaces, since it explores more candidates within the time limit due to fewer number of test executions. This explains the effectiveness and the efficiency of F1X compared with other tools shown in Table 2 and Figure 11. Recall that F1X$^E$ is a variant of F1X without test-equivalence partitioning. For the same search space, F1X$^E$ explores less candidates than F1X within the time limit since it requires more test executions. From this observation, we conclude that test-equivalence partitioning is responsible for the higher scalability of F1X.

**Table 3: Exploration statistics of program repair tools in stop-after-first-found and full exploration modes.**

| Subject | Stop-after-first-found mode | | | | | Full exploration mode | | | | |
|---|---|---|---|---|---|---|---|---|---|---|
| | F1X | F1X$^E$ | PR | PR* | GP | F1X | F1X$^E$ | PR | PR* | GP |
| libtiff | $\frac{44675}{243}$=183.8 | $\frac{1241}{1272}$=1.0 | $\frac{14980}{2946}$=5.1 | $\frac{6995}{4625}$=1.5 | $\frac{1371}{1302}$=1.1 | $\frac{400786}{759}$=528 | $\frac{23872}{24512}$=1.0 | $\frac{62521}{23763}$=2.6 | $\frac{31403}{16777}$=1.9 | $\frac{4564}{4657}$=1.0 |
| lighttpd | $\frac{1085}{220}$=4.9 | $\frac{512}{616}$=0.8 | $\frac{4645}{1295}$=3.6 | $\frac{6269}{2563}$=2.4 | $\frac{359}{383}$=0.9 | $\frac{112667}{850}$=132 | $\frac{10283}{10580}$=1.0 | $\frac{45338}{8377}$=5.4 | $\frac{29106}{10126}$=2.9 | $\frac{2337}{2371}$=1.0 |
| php | $\frac{2843}{1717}$=1.7 | $\frac{3407}{5184}$=0.7 | $\frac{1186}{48070}$=0.0 | $\frac{2566}{56492}$=0.0 | $\frac{88}{7736}$=0.0 | $\frac{77472}{193}$=401 | $\frac{9801}{12139}$=0.8 | $\frac{6378}{58671}$=0.1 | $\frac{4942}{73898}$=0.1 | $\frac{2057}{3497}$=0.6 |
| gmp | $\frac{5517}{173}$=31.9 | $\frac{30}{31}$=1.0 | $\frac{3215}{2002}$=1.6 | $\frac{2064}{1260}$=1.6 | $\frac{9128}{9172}$=1.0 | $\frac{80448}{1140}$=70 | $\frac{16934}{23994}$=0.7 | $\frac{17526}{14523}$=1.2 | $\frac{11423}{10156}$=1.1 | $\frac{2841}{2841}$=1.0 |
| gzip | $\frac{13071}{123}$=106.3 | $\frac{241}{251}$=1.0 | $\frac{5892}{1340}$=4.4 | $\frac{5284}{890}$=5.9 | $\frac{7605}{5735}$=1.3 | $\frac{518803}{568}$=913 | $\frac{18501}{24789}$=0.7 | $\frac{59467}{28716}$=2.1 | $\frac{48491}{40574}$=1.2 | $\frac{9007}{9098}$=1.0 |
| python | $\frac{940}{231}$=4.1 | $\frac{29}{37}$=0.8 | $\frac{4211}{7389}$=0.6 | $\frac{6928}{8771}$=0.8 | $\frac{3327}{3319}$=1.0 | $\frac{105842}{1617}$=65 | $\frac{8381}{12494}$=0.7 | $\frac{13367}{9213}$=1.5 | $\frac{9928}{9050}$=1.1 | $\frac{7823}{7824}$=1.0 |
| wshark | $\frac{828}{139}$=6.0 | $\frac{595}{597}$=1.0 | $\frac{3744}{2460}$=1.5 | $\frac{3948}{2434}$=1.6 | $\frac{4253}{4297}$=1.0 | $\frac{689925}{620}$=1112 | $\frac{14094}{14003}$=1.0 | $\frac{23043}{13099}$=1.8 | $\frac{18554}{17442}$=1.1 | $\frac{2008}{2213}$=0.9 |
| fbc | $\frac{948}{434}$=2.1 | $\frac{650}{891}$=0.7 | $\frac{1111}{672}$=1.65 | $\frac{524}{394}$=1.33 | $\frac{22}{788}$=0.0 | $\frac{50195}{1312}$=38 | $\frac{20195}{24521}$=0.8 | $\frac{892}{589}$=1.51 | $\frac{766}{499}$=1.54 | $\frac{852}{852}$=1.0 |

## 6.4 Effect of equivalent relation (RQ3)

Table 4 shows the effect of the three equivalent relations ($\overset{t}{\sim}_{value}$, $\overset{t}{\sim}_{deps}$, and $\overset{t}{\sim}_{*}$) in F1X on the average number of test executions for Libtiff. For each transformation from Figure 10, the table demonstrates the number of locations in which the transformation was applied (the "Locations" column), the test-equivalence relations that was applied for the search space produced by this transformation (the "Relation" column), the number of different candidate patches generated (the "Candidates" column), the number of partitions that were identified for the failing test (the "Partitions" column) and the number of tests required to explore all the corresponding candidates with the whole test suite (the "Test executions" column).

These results demonstrate that for the transformations $M_{EXPRESSION}$, $M_{REFINEMENT}$ and $M_{GUARD}$ our algorithm produces a small number of test-equivalence classes for the failing tests (with 500-1000 elements in each partition on average) which also resulted in a small number of executions for the whole test suite. For the relation $M_{ASSIGNMENT}$, the reduction is less significant, however the composition of relations $\overset{t}{\sim}_{*}$ is significantly more efficient than the individual relations $\overset{t}{\sim}_{value}$, $\overset{t}{\sim}_{deps}$.

## 7 RELATED WORK

The concept of test-equivalence have been previously utilized in several domains. Our work is the first that applies it to automatic patch generation and shows that it yields a substantial performance improvement compared with previous program repair techniques.

*Program synthesis.* Existing program synthesis techniques can be used to generate patches by directly searching in patch spaces, however this approach has limitations as explained in the following. First, the cost of test execution in a typical program repair problem is substantially higher than that in program synthesis, since the subjects on which program repair is carried out today are significantly larger (e.g. a single test execution for PHP interpreter from GenProg benchmark [Le Goues et al. 2012a] takes 1-10 seconds on commodity hardware, while a typical solution in SyGuS-Comp competition can be executed in $10^{-6}$ seconds). Second, the complexity of repaired programs makes it infeasible to apply precise deductive techniques. For instance, Sketch [Solar-Lezama 2008] fills "holes" in sketches (partial programs), and can be potentially applied to generate repairs for identified suspicious statements. However, it translates programs into boolean formulas and therefore can repair only relatively small programs [Hua and Khurshid 2016]. Since program synthesis algorithms may not be directly applicable to program repair, they are used as parts of program repair algorithms for filling "holes" in programs based on inferred specification [Long and Rinard 2015; Mechtaev et al. 2016]. In our technique, we do not use program synthesis as a black box, but integrate synthesis with program analysis by imposing additional requirements on the underlying synthesizer: support for the value-projection operator (Section 3.2). Since our implementation uses an enumerative program synthesis, it is straightforward to realize such operators. However, other techniques can also be used for this purpose. For instance, FlashMeta [Polozov and Gulwani 2015] compactly represents its search space as version space algebra (VSA) [Mitchell 1982]. Moreover, it defines the operation *Filter* over this representation that is effectively the value-projection operator, therefore it can be potentially used in our algorithm as an efficient representation of the space of program modifications.

**Table 4: Effect of equivalence relations on the number of test executions**

| Transformation | Relation | Locations | Candidates | Partitions | Test executions |
|---|---|---|---|---|---|
| $M_{EXPRESSION}$ | $\overset{t}{\sim}_{value}$ | 109 | 428641 | 424 | 759 |
| $M_{REFINEMENT}$ | $\overset{t}{\sim}_{value}$ | 45 | 48942 | 87 | 154 |
| $M_{GUARD}$ | $\overset{t}{\sim}_{value}$ | 87 | 106389 | 199 | 347 |
| $M_{ASSIGNMENT}$ | $\overset{t}{\sim}_{value}$ | 121 | 17181 | 3436 | 5983 |
| | $\overset{t}{\sim}_{deps}$ | 121 | 17181 | 1809 | 2308 |
| | $\overset{t}{\sim}_{*}$ | 121 | 17181 | 480 | 671 |

*Program repair search algorithms.* Syntax-based techniques generate patches by enumerating and testing syntactic changes. Since (1) repair tools have to explore large search spaces to address many classes of defects and (2) test execution has high cost for large real-world programs, they scale to relatively small search spaces. GenProg-AE [Weimer et al. 2013] eliminates redundant executions by identifying functionally-equivalent patches via lightweight analyses. Instead of functional equivalence, our technique applies test-equivalence, which is a weaker and therefore a more effective relation (produces larger equivalence classes). *Semantics-based* techniques split search into two phases. First, they localize suspicious statements and *infer specification* for the identified statements that captures the property of "passing the test suite". Such specification can be expressed as logical constraints [Nguyen et al. 2013] or *angelic values* [Long and Rinard 2015; Xuan et al. 2016]. Second, they apply off-the-shelf program synthesizers in order to modify the selected statements according to this specification. To infer specification, existing techniques perform path exploration by altering test executions. Since the number of execution paths in programs can be infinite, these methods are subject to the *path explosion problem.* For instance, Nopol [Xuan et al. 2016], SPR [Long and Rinard 2015] and Prophet [Long and Rinard 2016b] enumerate values of conditional expressions (which is a special case of path exploration) in order to find angelic values that enable the program to pass the failing test. As shown in Section 2.1, such techniques may perform a large number of redundant executions that are avoided by our algorithm. SemFix [Nguyen et al. 2013] and Angelix [Mechtaev et al. 2016] are semantics-based techniques relying on symbolic execution and SMT-based synthesis. Since our methodology does not use symbolic methods, it is orthogonal to SemFix and Angelix from that point of view of underlying analysis. Existing syntax and semantic-based techniques are limited to modifying side-effect free expressions. For instance, they can only generate new assignments by copying them from other parts of the program. Meanwhile, we demonstrate that test-equivalence can scale assignment synthesis using a combination of value- and dependency-based analyses.

*Program repair prioritization approaches.* In order to address the test over-fitting problem [Smith et al. 2015], various techniques have been proposed to prioritize patches that are more likely to be correct. For instance, DirectFix [Mechtaev et al. 2015] prioritizes candidate patches based on syntactic distance, Qlose [D'Antoni et al. 2016] prioritizes patches bases on semantic distance, Prophet [Long and Rinard 2016b] utilizes information learned from human patches, ACS [Xiong et al. 2017] prioritizes patches based on information mined from previous version and API documentation, S3 [Le et al. 2017] prioritizes patches based on a combination of syntactic and semantic properties. Our technique finds the best patch (the global optimum) in its search space according to an arbitrary cost function. Meanwhile, previous techniques applied to large real-world programs did not provide such guarantees. Techniques based on genetic programming [Arcuri and Yao 2008; Le Goues et al. 2012b] and random search [Qi et al. 2014] guarantee only a local optimum by definition. Semantics-based repair techniques may miss the global optimum, which is shown in Section 2.2.

*Program repair using types and formal specification.* Several approaches utilize temporal logic formulas [Jobstmann et al. 2005], contracts [Wei et al. 2010] and types [Reinking and Piskac 2015] to guide program repair. Our test-equivalence analysis can potentially optimize these approaches. Besides, our test-driven patch generation algorithm can be used in a counterexample-guided refinement loop [Alur et al. 2013] to repair programs based on given formal specification.

*Mutation testing.* The scalability of program repair is related to scalability of mutation testing, since mutation testing also evaluates a large number of program modifications. To address this problem in mutation testing, a common approach is to reduce the number of mutation operators (transformations) to avoid redundant executions [Mresa and Bottaci 1999]. This approach may not be suitable for program repair, because program repair search spaces have to be rich enough to enable generation of non-trivial human-like repairs. Test-equivalence have been applied to scale mutation testing. Mutant analysis by Just et al. [Just et al. 2014] performs a pre-pass that partitions mutants based on infected states, which can thought of as a variant of $\overset{t}{\sim}_{value}$ relation. More recent techniques [Ma and Kim 2016; Wang et al. 2017] extend this approach by performing more fine-grained partitioning. Techniques for mutation testing cannot be directly applied for program repair since program repair search spaces include significantly more complex modifications (e.g. synthesizing expressions for "holes" in programs, synthesizing assignment statements).

*Compiler testing.* Equivalence modulo inputs (EMI) [Le et al. 2014, 2015] (a variant of test-equivalence) was successfully applied to compiler testing. Specifically, for a compiler, a program and an input, it generates input-equivalent variants of the program by altering *unexecuted* statements and checks that these variants compiled by the compiler produce the same outputs. The main difference of our technique is that it analyzes test-equivalent change in *executed* statements. Therefore, our test-equivalence analyses might be used to increase the efficiency of EMI for compiler testing.

## 8 DISCUSSION AND FUTURE WORK

We envision the following design of a future *general-purpose program repair system* (a system that is able to address many kinds of defects in commodity software). This system will (1) implement a large number of transformations to address many kinds of defects, (2) implement test-equivalence analyses for these transformations to ensure scalability and (3) implement intelligent search space prioritization strategies over the patch space, to address the overfitting problem. This work is a step towards such a design. In our future works, we plan to investigate the following aspects.

*Test-equivalence relations.* The effectiveness of the relation $\overset{t}{\sim}_{value}$ (the size of test-equivalence classes it induces) may depend on the size of the output domain of modified expressions. Since it identifies equivalence based on concrete values, it may not be effective for expressions of large output domains (e.g. strings). In future, we plan to investigate a generalization of this relation that defines two changes to be test-equivalent iff they drive test execution along the same path. Such a relation will be computed using dynamic symbolic execution [Godefroid et al. 2005].

*Transformations.* Existing program repair systems provide very limited support for repairing function calls. We hypothesize that a generalization of the relation $\overset{t}{\sim}_{deps}$ may enable repair systems to extend search spaces by incorporating function call transformations in a scalable fashion.

*Current limitations.* Although we demonstrated that our approach based on test-equivalence significantly outperforms previous techniques, it has several limitations. The proposed analysis assumes deterministic test execution. Besides, the current algorithm is designed to synthesize single-line patches (involving a single modification). It can be potentially extended for multi-line modifications using the approach of Angelix [Mechtaev et al. 2016].

*Subject Programs.* The subjects for the evaluation (GenProg ICSE'12 benchmark [Le Goues et al. 2012a]) were previously used for evaluating related approaches [Long and Rinard 2015, 2016b; Mechtaev et al. 2016]. GenProg ICSE'12 benchmark was constructed to address generalizability concerns in evaluation of repair tools [Le Goues et al. 2012a]. Moreover, the test suites were independently augmented [Qi et al. 2015] to avoid generation of implausible patches. Nevertheless, a possible threat to validity is that our results may not generalize for other programs and defects.

## 9 CONCLUSION

We propose a methodology of automatic patch generation based on on-the-fly test-equivalence analysis. Identifying test-equivalence partitions of the space of patches enables our algorithm to drastically reduce the number of required test executions and therefore scale program repair to larger search spaces. Specifically, we propose two test-equivalence relations: based on runtime values and dynamic dependencies respectively. Our evaluation of eight real-world programs shows that the suggested algorithm searches through significantly more patch spaces and yet at least an order of magnitude faster than all existing patch generation systems.

## REFERENCES

Rajeev Alur, Rastislav Bodik, Garvit Juniwal, Milo MK Martin, Mukund Raghothaman, Sanjit A Seshia, Rishabh Singh, Armando Solar-Lezama, Emina Torlak, and Abhishek Udupa. 2013. Syntax-guided synthesis. In *FMCAD*. IEEE, 1–8.

Andrea Arcuri and Xin Yao. 2008. A novel co-evolutionary approach to automatic software bug fixing. In *CEC*. IEEE, 162–168.

Loris D'Antoni, Roopsha Samanta, and Rishabh Singh. 2016. Qlose: program repair with quantitative objectives. In *CAV*. Springer, 383–401.

Patrice Godefroid, Nils Klarlund, and Koushik Sen. 2005. DART: directed automated random testing. In *PLDI*. ACM, 213–223.

Jinru Hua and Sarfraz Khurshid. 2016. A sketching-based approach for debugging using test cases. In *ATVA*. Springer, 463–478.

Susmit Jha, Sumit Gulwani, Sanjit A Seshia, and Ashish Tiwari. 2010. Oracle-guided component-based program synthesis. In *ICSE*. 215–224.

Barbara Jobstmann, Andreas Griesmayer, and Roderick Bloem. 2005. Program repair as a game. In *CAV*. Springer, 226–238.

René Just, Michael D Ernst, and Gordon Fraser. 2014. Efficient mutation analysis by propagating and partitioning infected execution states. In *ISSTA*. ACM, 315–326.

Vu Le, Mehrdad Afshari, and Zhendong Su. 2014. Compiler validation via equivalence modulo inputs. In *PLDI*. ACM, 216–226.

Vu Le, Chengnian Sun, and Zhendong Su. 2015. Finding deep compiler bugs via guided stochastic program mutation. In *OOPSLA*. ACM, 386–399.

Xuan Bach Dinh Le, Duc Hiep Chu, David Lo, Claire Le Goues, and Willem Visser. 2017. S3: syntax- and semantic-guided repair synthesis via programming by example. In *FSE*. ACM.

Claire Le Goues, Michael Dewey-Vogt, Stephanie Forrest, and Westley Weimer. 2012a. A systematic study of automated program repair: Fixing 55 out of 105 bugs for 8 each. In *ICSE*. IEEE, 3–13.

Claire Le Goues, ThanhVu Nguyen, Stephanie Forrest, and Westley Weimer. 2012b. Genprog: a generic method for automatic software repair. *TSE* 38, 1 (2012), 54–72.

Fan Long and Martin Rinard. 2015. Staged program repair with condition synthesis. In *FSE*. ACM, 166–178.

Fan Long and Martin Rinard. 2016a. An analysis of the search spaces for generate and validate patch generation systems. In *ICSE*. ACM, 702–713.

Fan Long and Martin Rinard. 2016b. Automatic patch generation by learning correct code. In *POPL*. ACM, 298–312.

Chi-Keung Luk, Robert Cohn, Robert Muth, Harish Patil, Artur Klauser, Geoff Lowney, Steven Wallace, Vijay Janapa Reddi, and Kim Hazelwood. 2005. Pin: building customized program analysis tools with dynamic instrumentation. In *PLDI*. ACM, 190–200.

Yu-Seung Ma and Sang-Woon Kim. 2016. Mutation testing cost reduction by clustering overlapped mutants. *Journal of Systems and Software* 115 (2016), 18–30.

Sergey Mechtaev, Jooyong Yi, and Abhik Roychoudhury. 2015. Directfix: looking for simple program repairs. In *ICSE*. IEEE, 448–458.

Sergey Mechtaev, Jooyong Yi, and Abhik Roychoudhury. 2016. Angelix: Scalable Multiline Program Patch Synthesis via Symbolic Analysis. In *ICSE*.

Tom M Mitchell. 1982. Generalization as search. *Artificial intelligence* 18, 2 (1982), 203–226.

Elfurjani S Mresa and Leonardo Bottaci. 1999. Efficiency of mutation operators and selective mutation strategies: An empirical study. *STVR* 9, 4 (1999), 205–232.

Hoang Duong Thien Nguyen, Dawei Qi, Abhik Roychoudhury, and Satish Chandra. 2013. SemFix: program repair via semantic analysis. In *ICSE*. 772–781.

Oleksandr Polozov and Sumit Gulwani. 2015. FlashMeta: a framework for inductive program synthesis. *OOPSLA* (2015), 107–126.

Yuhua Qi, Xiaoguang Mao, Yan Lei, Ziying Dai, and Chengsong Wang. 2014. The strength of random search on automated program repair. In *ICSE*. ACM, 254–265.

Zichao Qi, Fan Long, Sara Achour, and Martin Rinard. 2015. An analysis of patch plausibility and correctness for generate-and-validate patch generation systems. In *ISSTA*. ACM, 24–36.

Alex Reinking and Ruzica Piskac. 2015. A type-directed approach to program repair.. In *CAV (1)*. 511–517.

Reudismam Rolim, Gustavo Soares, Loris D'Antoni, Oleksandr Polozov, Sumit Gulwani, Rohit Gheyi, Ryo Suzuki, and Björn Hartmann. 2017. Learning syntactic program transformations from examples. In *ICSE*. IEEE Press, 404–415.

David Shriver, Sebastian Elbaum, and Kathryn T Stolee. 2017. At the end of synthesis: narrowing program candidates. In *ICSE NIER*. IEEE Press, 19–22.

Edward K Smith, Earl T Barr, Claire Le Goues, and Yuriy Brun. 2015. Is the cure worse than the disease? overfitting in automated program repair. In *FSE*. ACM, 532–543.

Armando Solar-Lezama. 2008. *Program synthesis by sketching*. University of California, Berkeley.

Shin Hwei Tan, Hiroaki Yoshida, Mukul R Prasad, and Abhik Roychoudhury. 2016. Anti-patterns in search-based program repair. In *FSE*. ACM, 727–738.

Yida Tao, Jindae Kim, Sunghun Kim, and Chang Xu. 2014. Automatically generated patches as debugging aids: a human study. In *Proceedings of the 22nd ACM SIGSOFT International Symposium on Foundations of Software Engineering*. ACM, 64–74.

Bo Wang, Yingfei Xiong, Yangqingwei Shi, Lu Zhang, and Dan Hao. 2017. Faster mutation analysis via equivalence modulo states. *ISSTA* (2017).

Yi Wei, Yu Pei, Carlo A Furia, Lucas S Silva, Stefan Buchholz, Bertrand Meyer, and Andreas Zeller. 2010. Automated fixing of programs with contracts. In *ISSTA*. ACM, 61–72.

Westley Weimer, Zachary P Fry, and Stephanie Forrest. 2013. Leveraging program equivalence for adaptive program repair: Models and first results. In *ASE*. IEEE, 356–366.

Yingfei Xiong, Jie Wang, Runfa Yan, Jiachen Zhang, Shi Han, Gang Huang, and Lu Zhang. 2017. Precise condition synthesis for program repair. In *ICSE*. IEEE Press, 416–426.

Jifeng Xuan, Matias Martinez, Favio Demarco, Maxime ClÃĺment, Sebastian Lamelas, Thomas Durieux, Daniel Le Berre, and Martin Monperrus. 2016. Nopol: Automatic Repair of Conditional Statement Bugs in Java Programs. *TSE* (2016).

Jooyong Yi, Umair Ahmed, Amey Karkare, Shin Hwei Tan, and Abhik Roychoudhury. 2017. A feasibility study of using automated program repair for introductory programming assignments. In *FSE*. ACM.